\newcommand{\be}{\begin{equation}}
\newcommand{\bea}{\begin{eqnarray}}
\newcommand{\ee}{\end{equation}}
\newcommand{\eea}{\end{eqnarray}}

\newcommand{\qa}{\alpha}
\newcommand{\qb}{\beta}
\newcommand{\qg}{\gamma}
\newcommand{\qG}{\Gamma}
\newcommand{\qd}{\delta}

\newcommand{\qe}{\varepsilon}

\newcommand{\qh}{\eta}

\newcommand{\qy}{\theta}

\newcommand{\qr}{\rho}
\newcommand{\qs}{\sigma}

\newcommand{\qt}{\tau}
\newcommand{\qf}{\varphi}
\newcommand{\qF}{\Phi}

\newcommand{\qO}{\Omega}

\newcommand{\arctg}{{\rm arctg}}

\newcommand{\sgn}{{\rm sgn}}

\newcommand{\tr}{{\rm tr}\;}
\newcommand{\Tr}{{\rm Tr}\;}

\newcommand{\inv}{^{-1}}

\newcommand{\dagg}{^{\dag}}
\newcommand{\prt}{\partial}

\newcommand{\intd}[1]{\int \!\! d{#1} \;}

\newcommand{\ointdx}{\oint\! dx\;}
\newcommand{\intdxx}{\int \!\! d^{2}x \;}

\newcommand{\fr}[2]{{\textstyle \frac{#1}{#2}}}
\newcommand{\half}{\mbox{$\textstyle \frac{1}{2}$}}

\newcommand{\naar}{\rightarrow}

\newcommand{\nn}{\nonumber}

\newcommand{\scez}{\setcounter{equation}{0}}
\newcommand{\ns}{\scez\section}

\renewcommand{\theequation}{\thesection .\arabic{equation}}

\newcommand{\dn}{\triangle\nu}
\newcommand{\vd}{v_{\rm d}}
\newcommand{\veff}{v^{\rm eff}}
\newcommand{\eff}{^{\rm eff}}

\newcommand{\Ian}{{\rm I}^\qa_n}
\newcommand{\Iamn}{{\rm I}^\qa_{-n}}

\newcommand{\curl}{\nabla\!\times\!}

\documentstyle[preprint,aps,eqsecnum,epsf,prb, tighten]{revtex}
\begin{document}

\draft

\title{The fractional quantum Hall effect: Chern-Simons mapping,
duality, Luttinger liquids and the instanton vacuum}

\author{B. \v{S}kori\'{c}$^*$ and A.M.M. Pruisken$^{*\dagger}$}

\address{$^*$Institute for Theoretical Physics, University of Amsterdam,
Valckenierstraat 65, \\ 1018 XE Amsterdam, The Netherlands}

\address{$\dagg$Institute for Theoretical Physics, University of California,
Santa Barbara, CA 93106-4060}

\maketitle

\begin{abstract}
\noindent
We derive, from first principles, the complete Luttinger liquid theory
of abelian quantum Hall edge states. This theory includes the effects
of disorder and Coulomb interactions as well as the coupling to external
electromagnetic fields. We introduce a theory of spatially separated
(individually conserved) edge modes, find an
enlarged dual symmetry and obtain a complete classification
of quasiparticle operators and tunneling exponents.
The chiral anomaly on the edge and Laughlin's gauge argument are used to obtain
unambiguously the Hall conductance. 
In resolving the problem
of counter flowing edge modes, we find that the long range
Coulomb interactions play a fundamental role.
In order to set up a theory for arbitrary filling fractions $\nu$ we
use the idea 
of a two dimensional network of percolating edge modes. We derive an effective,
single mode Luttinger liquid theory for tunneling processes into the 
quantum Hall edge which yields a continuous tunneling exponent $1/\nu$.
The network approach is also used to re-derive the
instanton vacuum  
or $Q$-theory for the plateau transitions.

\end{abstract}

\newpage

\tableofcontents


\ns{Introduction}

Many of the unusual aspects of both integral and fractional quantum
Hall states can be understood  from the properties of the excitations
at the sample edge. Effective theories for edge excitations take the
form of a chiral Fermi liquid in the integral regime or a chiral
Luttinger liquid\cite{Wen}
in the fractional regime. These theories are simple
examples of so-called conformal field theories that are commonly
used as a computational scheme for physical processes such as
tunneling into the quantum Hall edge 
states.\cite{KFtunnel,Wentunnel,Chang,continuum} 
They also provide an interesting model system for exploring the thermodynamic
consequences of quasiparticle excitations with unusual statistics
(exclusion statistics).\cite{exclusion}

The standard Luttinger liquid approach to the quantum Hall edge has
remained largely phenomenological and ad hoc in nature. It reflects
the idealized properties of "pure" Laughlin states\cite{Laughlin}
that are
characterized by an {\em energy gap} and therefore, the analysis is
limited to a set of special filling fractions only. 
But even within its limited range of validity it is not clear
how to deal with the fundamental aspects of disorder. For example,  
controversial issues have arisen regarding the Hall conductance of
more complicated states that have edge channels of opposite
chirality.\cite{KaneFisher,KFP} 
Moreover, the lack of a microscopic theory of the fqHe has
led to serious discrepancies between
experiments on edge tunneling\cite{continuum} 
on the one hand and the Luttinger liquid
approach to the quantum Hall edge on the other.  

One of the main objectives of the present work is to derive, from
first principles,  the complete Luttinger liquid theory for abelian
quantum Hall states.  
In our approach to the problem, the physics of the edge
appears as an exactly solvable, {\em strong coupling} limit of a more
general theory for arbitrary filling fractions that deals simultaneously with 
disorder effects and the Coulomb interactions. In a recent series of
articles\cite{Mishandling1,Mishandling2,Mishandling3},
hereafter referred to as [I], [II] and [III],
we have laid  the foundation for 
such a unifying theory. The basic starting point is the Finkelstein
sigma model theory ($Q$ theory) for localization and  
interaction effects.\cite{Finkelstein} 
The fundamental principle underlying the unification of compressible
and incompressible quantum Hall states is the so-called {\em instanton
vacuum}, 
a topological concept in quantum field theory that was discovered
earlier by one of us in the context of the {\em free} electron
localization theory.\cite{thetavac} 

In [III] we established the relation between the {\em instanton vacuum } 
or $Q$-field approach to the iqHe and
and the theory of {\em chiral edge bosons}.
In this paper we build on these results
and extend the theory to include the fractional effect.
For this purpose, we shall employ the idea of flux attachment\cite{Jain} 
by coupling the $Q$-field theory to the Chern-Simons
statistical gauge fields.\cite{DJT,Schonfeld}
Flux attachment transformations have been applied 
several times before and with different physical 
objectives.\cite{Fradkinbook,Zhang92,HLR,WenZee92,Ezawa,LF95}
A new ingredient of the present work is that our
detailed knowledge of the {\em integral} quantum Hall
edge can now be "mapped" directly onto a complete Luttinger liquid
theory of the {\em fractional} quantum Hall edge. 
Mathematically, this "mapping" is accomplished by integrating out the
Chern-Simons gauge fields over the entire
two dimensional plane while confining the electronic degrees of
freedom to the half plane. This procedure then leads to
the familiar $K$-matrix structure of the edge 
excitations.\cite{FrohlichZee,FrohlichKerler,FrohlichStuder,BlokWen1990,Wen} 
As a major advancement, however, we shall point out that the 
flux attachment procedure also provides a simple
geometrical interpretation of {\em duality} 
transformations\cite{Balachandran} 
that invert the $K$-matrix (compactification radius) of  
the Luttinger liquid edge excitations. 

In order to facilitate an analysis of slowly 
varying potential fluctuations, we introduced in [III] the concept of 
"spatially separated edge
modes". We argued that smooth potential fluctuations near the sample edge
decompose the multiple (iqHe) edge states into physically
distinguishable and separately conserved edge modes. 
Since in our theory of the dirty, sharp edge the various edge modes are
assumed to scatter strongly amongst each other, it is a priori not 
entirely obvious
whether the concept of spatially separated edge modes "maps" onto 
a physically
equivalent fractionally quantum Hall state under the flux attachment 
transformation.
This problem is closely related to the problem of {\em counter flowing
edge modes} and  
to a number of puzzles\cite{LopezFradkin,LeeWen} that recently arose in
the context of the conventional $K$-matrix  approach.

By extending the flux attachment transformation to include the case
of widely separated edge modes 
we obtain a rich structure of decomposed fractional quantum Hall edge states
which is reminiscent of the picture of multiple condensates in the hierarchy 
theory of the fqHe.\cite{Read,Haldane83,Halperin84} 
The Luttinger liquid theory for spatially separated edge modes
has quite interesting pictorial aspects. In this case we
find an {\em enlarged dual symmetry} of 
abelian quantum Hall states. We employ the enlarged dual 
symmetry in order to derive a complete classification of operators
for fractionally charged edge excitations as well as tunneling amplitudes.

In the limit of long wavelengths (i.e. for distances large compared to the 
spatial separation between the edge modes)
our theory of separated edge modes reduces in effect to a Luttinger liquid 
theory of a sharp, clean edge.
We next use our explicit results for the "clean" and "dirty" edges to show 
that they describe identical physics
in the limit of large length scales. This not only means that the fractional 
quantum hall {\em conductance} is the 
same in both cases,
but also identical amplitudes are obtained for the tunneling of electrons and 
quasiparticles. 
For incompressible states with only chiral edge modes these results are 
precisely what one would naively expect.
For more complicated states with both chiral and anti-chiral
components present, however, our theory is 
quite different from the approaches that have been pursued previously. 

In view of the fact that the Luttinger liquid theory of the quantum Hall 
edge has already been extensively investigated
following the phenomenological 
theory of Wen\cite{Wen}, it may be helpful to stress some of the main
differences between the standard approach and the  
development of a microscopic theory of the edge as pursued in this paper.

First, our edge theory is coupled to the external electromagnetic 
field throughout the derivation. This obviously has considerable 
technical as well as conceptual advantages. For example, the appearance 
of the so-called {\em chiral anomaly}\cite{Balachandran,Maeda}
plays a crucial role in relating
the Hall conductance as an {\em edge} phenomenon 
unambiguously
to the Hall conductance as a {\em bulk} phenomenon. 
In this way we avoid mistakes in applying ad hoc pictures 
like the Landauer-B\"{u}ttiker formalism.\cite{KaneFisher}
The chiral anomaly also provides an explicit demonstration of Laughlin's
gauge argument\cite{PrangeGirvin} 
and it plays an important role in identifying the
fractionally charged edge excitations in more complex situations.

Secondly, the Chern-Simons mapping or flux attachment procedure is
ill-defined in 
the absence of long range interactions between the particles. For example,
without the Coulomb interactions being present, the Hamiltonian 
for counter flowing edge modes would become
unbounded from below. The fundamental significance of long range interactions
is necessarily overlooked in the phenomenological approach, where
one is free to choose boundary
conditions (velocities) and bounded Hamiltonians even for
noninteracting particles. The fact that long range Coulomb
interactions are needed for a proper CS transformation is related to
the central role that they play in the 
${\cal F}$-invariance of the $Q$-field theory.

Thirdly, there is a difference in the way that quasiparticle operators and 
tunneling exponents are
obtained. We shall show that the flux attachment procedure leaves no room for
ambiguities, since it directly maps the electron operators of the
integral quantum  
Hall edge into quasiparticle operators of the fractional quantum Hall edge.
This procedure discriminates between opposite sides of an edge, thus giving 
a novel, geometrical interpretation to fractionally charge edge excitations. 
Physically this means that tunneling processes are described by
different particles 
with different charge and statistics, depending on whether one
tunnels, say, through the vacuum  
or through the incompressible quantum Hall state. 

Finally, our velocity matrix is diagonal simultaneously with $K$.
Since impurities have been dealt with in the first stages of our
derivation,  they do not make a second appearance in the form of
inter-channel scattering of chiral bosons. Consequently, the
calculation of quasiparticle and tunneling operators is not plagued
by the kind of 
non-universalities found by\cite{KaneFisher,KFP} for
counter flowing charged and neutral modes.

\vskip4mm

In the second part of this paper we embark on the more general problem
of {\em arbitrary}  filling fractions in the fractional quantum Hall
regime. We mentioned earlier that our 
Luttinger liquid theory of edge modes merely describes the strong
coupling asymptotics of a unifying theory for arbitrary filling
fractions. A complete understanding of this theory 
is beyond the scope of the present work since it involves a general
scaling scenario between  extreme physical states that can rarely be
followed in a single experiment. If one limits  
oneself to realistic quantum Hall conditions then a two dimensional,
percolating network of  edge modes presumably is the appropriate
starting point for an analysis of continuously  
varying filling fractions. Our theory of spatially separated edge
modes enables us to formulate such a percolating network, similar to
what was introduced previously in the context of 
the integral quantum Hall regime. In fact, most of the analysis
previously done for integral quantum Hall 
states can be directly "mapped", under the flux attachment
transformation, onto the  fractional quantum Hall regime.  More
specifically, the results obtained for the filling fractions  varying
between $m$ and $m\!+\!1$ in the integral regime transform, under 
an attachment of $2p$ flux quanta per electron, into equivalent
statements made on filling fractions varying between $\nu_m$ and
$\nu_{m+1}$ in the fractional regime. Here, $\nu_m \!=\!m/(2pm\!+\!1)$ 
denotes the Jain series.  
Our percolating network of edge modes explains two fundamentally
distinct aspects of the fractional 
quantum Hall regime that cannot be obtained from the theory of
Laughlin states or isolated 
edge excitations alone. These fundamental aspects are easily
recognized experimentally. In particular, 
the {\em plateau features} that are usually observed on the Hall
conductance with varying filling fraction $\nu$ are
clearly not seen in the recent experiments on electron 
tunneling\cite{continuum} 
into the quantum Hall edge.  
The presence or absence of plateau features with varying $\nu$
clearly demonstrates the fundamental differences 
in transport and equilibrium properties of the electron gas. 

Following [III] we use the idea of percolating
edge states and derive an "effective" Luttinger liquid theory for
tunneling processes into the fractional 
quantum Hall edge. Here, the main idea is that the long range Coulomb
interactions between the edge and  
the localized bulk orbitals give rise to an effective, chiral action
of edge excitations in which the 
neutral modes are strongly suppressed. Only the charged mode
contributes to the tunneling processes and 
the effective Luttinger liquid parameter varies continuously with the
filling fraction $\nu$. 
This yields a tunneling exponent $1/\nu$ in
accordance with experiments.

The plateau features in the Hall conductance are explained by the
localization of quasiparticles in the
bulk and this demands a totally different theory of {\em plateau transitions}.
We shall employ the percolating network model and briefly derive the
$Q$-field theory for localization and interaction effects. The results
indicate that the critical aspects
of the plateau transitions are quite generally the same for both the
integral and fractional regime, 
independent of the type of disorder (long range vs. short
range). However, our results do not exhibit 
the Sl(2,Z) symmetry that has been proposed and advocated 
in\cite{LR93,Lutken93,KLZ92}.
This symmetry, therefore,  
has no microscopic justification and it does not appear as a
fundamental (dual) symmetry of the quantum phase transition.

The organization of this paper is as follows.
In Section~\ref{secchirbos} we describe the details of the flux
attachment transformation 
for the disordered edge. The $Q$-field or instanton vacuum representation and
the equivalent chiral boson theory for the integral quantum Hall regime 
are reviewed in Section~\ref{secbosint}. 
In Section~\ref{secthefrac} 
we introduce the Chern Simons statistical gauge fields and,
by integrating 
them out, obtain the complete theory for fractional quantum Hall edges.
The most important results are summarized by the boson actions of Eqs.
(\ref{mappedch2}) and (\ref{basisneutral}).
These include the effects of the Coulomb interaction as well as the
coupling of  
the theory to external electromagnetic fields. Some comments on the problem
of counter flowing  edge modes are presented in 
Section~\ref{secinstability}.

In section~\ref{secshorttunn} we extend the Chern Simons mapping 
of the disordered edge to include the electron operators and tunneling
exponents. We reproduce the Kane-Fisher-Polchinski results\cite{KFP} 
for the tunneling exponents and show the geometrical structure of duality.

In Section~\ref{seclongrange} we introduce the concept of {\em
spatially separated edge channels} 
and repeat the various steps of flux attachment. A complete set of
tunneling exponents is 
given in (\ref{isolatedS}).
Section~\ref{secdual} deals with the subject of
{\em enlarged} duality. 
In Section~\ref{secPhiphi} we compare the Luttinger liquid theory of spatially
separated edge modes with theory  
of the dirty edge. In Section~\ref{quasiLaughlin} 
we derive the operators for the  Laughlin quasiparticles.

In Section~\ref{seclongtunn} we apply the idea of percolating edge
modes and derive an 
effective Luttinger liquid theory 
for tunneling processes away from the special filling fractions.

Section~\ref{secPlattrans} is devoted to the $Q$-theory of
the plateau transitions 
and in Section~\ref{secsummary} we conclude with a  summary of the results.


\ns{FQHE chiral bosons on a sharp edge}
\label{secchirbos}

\subsection{Interacting chiral edge bosons for the integer effect}
\label{secbosint}

\subsubsection{$Q$ fields at the edge}

In [I-III] we derived the Finkelstein sigma model ($Q$ field) action
for disordered,  
interacting 2D electrons in a strong magnetic field.
Specializing to the case where the chemical potential $\mu$ lies in an
{\em energy gap} between two Landau bands, the action reduces to a 
theory for {\em edge excitations} alone. 
A detailed understanding of the dynamics of edge excitations 
is important since it provides invaluable information on the 
topological concept of an {\em instanton vacuum} in strong coupling. 

In [III] we presented a detailed derivation
of the complete "edge" theory coupled to an external electromagnetic
field $A_\mu$. 
Since we are dealing with an exactly soluble, strong coupling limit of an
otherwise extremely non-trivial theory we shall proceed by summarizing
the results. 
We refer to [I-III] for a detailed discussion on symmetries, gauge
invariance etc.  
In terms of the
matrix field variables $Q^{\qa\qb}_{nm}$ the edge theory can be
expressed as follows 
\bea
        S_{\rm eff}[Q,A]&=&m S_{\rm top}[Q]+\fr{im}{4\pi}\left[
        \sum_\qa\int\!(A\eff)^\qa\wedge d(A\eff)^\qa-\qb\sum_{n\qa}
        \oint\! dx\; (A_x)^\qa_{-n}(A_c\eff)^\qa_n\right] \nn\\
        &&+\fr{m\pi}{4\qb\vd}S_{\rm F}[Q]-\fr{m\pi}{4\qb}\sum_{n\qa}
        \int\! \fr{d k_x}{\veff(k_x)}
        \left[\tr\Iamn Q-\fr{\qb}{\pi}(A_c\eff)^\qa_n\right]
        \left[\tr\Ian Q-\fr{\qb}{\pi}(A_c\eff)^\qa_{-n}\right] \nn\\
        && +\fr{\qb}{2}(\fr{m}{2\pi})^2\sum_{n\qa}\int\! d^2 x d^2 x'\;
        B^\qa_n(\vec x) U(\vec x-\vec x') B^\qa_{-n}(\vec x') .
\label{Qform}
\eea
Here, $m$ denotes the integer filling fraction and  $S_{\rm top}$
stands for the topological term
\be
        S_{\rm top}[Q]=\fr{1}{8}\intdxx \qe_{ij}\tr Q\prt_i Q\prt_j Q
\ee
which can also be written as an integral over the sample edge. The
$S_{\rm F}$ stands for the one dimensional Finkelstein action
\be
        S_{\rm F}[Q]=\oint\! dx \left[\sum_{n\qa}\tr\Ian Q\tr \Iamn Q 
        +4\tr\qh Q\right].
\ee
The explanation of the remaining symbols is as follows.
The indices $\qa , \qb$ denote replica channels, the $n$ are Matsubara
frequency indices and $B$ is the
magnetic field. The inverse temperature $(k_B T)\inv$ is written as
$\qb$. The matrix $\qh$ is given by
$\qh^{\qa\qb}_{kl}\!=\!\qd^{\qa\qb}\qd_{kl}k$.

We have taken the line $y\!=\!0$ as the
sample's edge and the half-plane $y\!>\!0$ as the bulk.
The integral $\oint\!dx$ denotes integration over the edge and 
$\intdxx$ should be read as an integral over the upper half-plane.
The integral containing the ``wedge'' notation is defined according to
\be
        \int\! A\wedge dA := \int_0^\qb\!\! d\qt \int_{-\infty}^\infty
        \!\! dx \int_0^\infty\!\! dy\; \qe^{\qa\qb\qg}A_\qa\prt_\qb A_\qg,
\label{CSnotation}
\ee
where the coordinate $x^0$ denotes the imaginary time $\qt$.
All the fields that we work with in this paper have bosonic (periodic)
boundary 
conditions in the imaginary time coordinate $\qt$.
We work in units where $\hbar\!=\!e\!=\!1$, and all lengths are
expressed in terms of the magnetic length 
$\ell \!=\! \sqrt{2\hbar/ eB}$.
The superscript $\eff$ indicates that the scalar potential $A_0$ has
acquired a contribution from the Coulomb interactions,
\bea
        A_j\eff=A_j \hskip5mm &;& \hskip5mm
        A_0^{\rm eff}(\vec{x})= A_0(\vec{x})
        +\fr{im}{2\pi}\intd{^2 x'}
        U(\vec{x}-\vec{x}')B(\vec{x}').
\label{defA0eff}
\eea
The subscript `$c$' denotes a chiral direction with an extra
interaction term,
\bea
        A_c=A_0-iv\eff A_x  \hskip5mm &;& \hskip5mm
        \prt_c=\prt_0-iv\eff\prt_x \nn\\
        \veff(k_x) = \vd+\fr{m}{\sqrt{2\pi}}U(k_x) \hskip5mm &;& \hskip5mm
        \vd=m/(2\pi\qr_{\rm edge})
\eea
where $\qr_{\rm edge}$ is the density of states at the edge.
The $U(\vec x\!-\!\vec x')$ is the Coulomb interaction, proportional to
$|\vec x\!-\!\vec x'|\inv$. The restriction of this interaction to the edge
is given (in momentum space) by 
$U(k_x)\!=\!\fr{1}{\sqrt{2\pi}}\intd{k_y}U(\vec k)$.

\subsubsection{Chiral edge bosons}

After these preliminaries we quote a completely equivalent but much simpler
theory for edge dynamics which is expressed in terms of chiral edge
boson fields $\phi_i$ 
with the index $i$ running from $1$ to $|m|$. In [III] we obtained the
complete chiral boson  action as follows
\bea
\label{Cbosonform}
        S_{\rm chiral} &=& \fr{i}{4\pi}\sgn(m)
        \sum_{i=1}^{|m|}\left[
        \int\!\! A\wedge dA
        -\oint(D_x\qf_i D_-\qf_i-E_x \qf_i)
        \right] \\ &&
        -\fr{1}{2}(\fr{1}{2\pi})^2\sum_{i,j=1}^{|m|}
        \int_\infty\!\! d\qt d^2 x d^2 x'\;
        \; U(\vec x-\vec x')\;\curl[\qy(y)\vec{D}\qf_i(\vec x)]
        \; \nabla'\!\times\![\qy(y')\vec{D}\qf_j(\vec x')] . \nn
\eea
Here the covariant derivative is defined as
$D_\mu\qf\!=\!\prt_\mu\qf \!-\! A_\mu$.
The  $\qy$ is the Heaviside step function and 
$D_-\qf\!=\!D_0\qf\!-\!i\vd D_x\qf$.
We use the shorthand notation $\oint$ for 
$\int_0^\qb\! d\qt\ointdx$.
Notice that 
there are no `effective' quantities in (\ref{Cbosonform});
The Coulomb interaction is completely contained in the last term.
The charge density is given by
$\fr{m}{2\pi}[B+\qd(y)|m|\inv\sum_i D_x\qf_i]$.
Notice also that we have written a two-dimensional integral containing
$\qf_i$, even though the boson fields only exist on the edge. 
This is not a problem, since the $\qf_i$ only get evaluated at the
edge.

The equivalent theories of edge dynamics, (\ref{Qform}) 
and (\ref{Cbosonform}), provide an important
tool for investigating the  problem of "long range potential fluctuations"
in quantum Hall systems. For example, in [III] we introduced a 2D network of
percolating edge modes as a model for long range potentials and
arbitrary filling  
fractions $\nu$. It was shown that tunneling processes at the quantum Hall edge
are conveniently described by an {\em effective} theory of chiral edge bosons.
The physics of the plateau transitions, on the other hand, is
completely different 
and the original instanton vacuum or $Q$ theory can be re-derived from
the network model,  
starting from (\ref{Qform}).

The Chern Simons mapping, described below, proceeds entirely in terms of the 
chiral boson version (\ref{Cbosonform}). 
We return to the $Q$ fields in Section~\ref{secPlattrans} where we  
embark on the subject of the plateau transitions.

\subsection{The fractional effect}
\label{secthefrac}

\subsubsection{Composite fermions}

In writing (\ref{Cbosonform}), we have included the
possibility of $m$ being negative. The reason for this is that we want
to be able to perform Jain's composite fermion mapping \cite{Jain},
which relates integer
filling $m$ to fractional values $\nu$, in such a way that the electron and
the composite fermion experience magnetic fields of opposite sign.
We use a convention in which 
$\nu$ is always positive and the magnetic field of the filling
fraction $\nu$ sample points in the positive $z$ direction.
Since particle
densities are positive, we then have to allow negative $m$. 

We implement the composite fermion picture in the standard way 
by introducing statistical gauge fields $a_\mu$. The action for the
$a_\mu$ has the form of a Chern-Simons
action\cite{DJT,Schonfeld} and our starting point can be written as
\be
\label{SCS2}
        S = S_{\rm chiral}[\qf_i,A+a]+
        \fr{i}{4\pi}\cdot\fr{1}{2p}
        \int_\infty\!\!\! a \wedge d a
\ee
with $S_{\rm chiral}$ given by (\ref{Cbosonform}).
Here $p$ is a positive integer and the subscript $\infty$ denotes
that the spatial integral is  
over the whole $xy$-plane instead of just the upper half.

\begin{figure}
\begin{center}
\setlength{\unitlength}{1mm}
\begin{picture}(70,40)(0,0)
\put(0,5)
{\epsfxsize=70mm{\epsffile{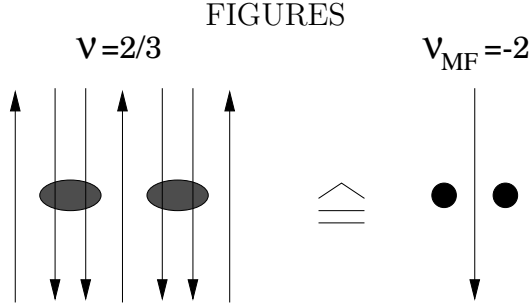}}}
\end{picture}
\caption{Equivalence between filling fraction $2/3$ of electrons
coupled to a CS gauge field (left) and filling $-2$
in mean field theory (right). Arrows denote flux quanta. The dots
represent electrons.}
\label{figtwothirds}
\end{center}
\end{figure}

Several remarks should be added to this result.
First of all, the theory of (\ref{SCS2}) does not contain the zero frequency
modes and it really stands for the {\em fluctuations} about the
Chern Simons mean field theory. As a result, the
$S_{\rm chiral}$ in (\ref{SCS2}) is defined for an effective magnetic field
$B\eff$ resulting from a ``smearing out'' of the
statistical fluxes. It is given by [I]
\be 
        B\eff\!=\!B\!-\! 2p n_e, 
\label{Beff}
\ee
with $n_e$ the electron density. The $2p$, on a mean field level, stands
for the number of elementary
flux quanta attached to each electron. 
The only effect of (\ref{SCS2}) is that the integer filling fraction $m$
now refers to {\em composite fermions} rather than electrons.
The true filling fraction $\nu$ of the electron gas
is related to $m$ in the following way,
\be
        \nu=\frac{m}{2pm+1},
\label{nu(m,p)}
\ee
where $m$ can be negative if the signs of $B$ and $B\eff$ are not the
same. 
As an illustration we show in Fig.~\ref{figtwothirds} that the
$\nu\!=\!2/3$ state 
is equivalent to integer filling $m\!=\!-2$ at a mean field level if
two flux quanta per electron are applied.

Secondly, it is important to stress that the statistical gauge fields
$a_{\mu}$ are 
defined over the entire plane. The electronic part $S_{\rm chiral}$ is
restricted to the half plane 
and this, properly regarded, is the result of a confining potential
added to the electron gas.

Keeping these preliminary remarks in mind we next turn to the problem
of the Chern Simons fluctuating gauge fields.

\subsubsection{Integrating out the CS fluctuations}

We note that the action can be written in such a way that
$a_-(=\!a_0\!-\!i\vd a_x)$, the component in the ``preferred'' chiral
direction, 
evidently multiplies a constraint,
\bea
\label{multform}
	S &=& \fr{i}{4\pi}\left[\fr{1}{2p}\int_\infty(2a_-\curl\vec{a}
	-\vec{a}\!\times\!\prt_-\vec{a}) \right. \\ && \left.
	+\sgn(m)\sum_{i=1}^{|m|}\int_\infty
	\left(2{\cal D}_-\qf_i\curl
	[\qy(y)\vec{\cal D}\qf_i]-\qy(y)\vec{\cal D}\qf_i\!\times\!\prt_-
	\vec{\cal D}\qf_i\right)
	\right] \nn\\ &&
	-\fr{1}{2}(\fr{1}{2\pi})^2\sum_{i,j=1}^{|m|}
	\int_\infty U(\vec x-\vec x')\;
	\curl[\qy(y)\vec{\cal D}\qf_i(\vec{x})]\;
	\nabla'\!\times\![\qy(y')\vec{\cal D}\qf_j(\vec{x}')].\nn
\eea
Here ${\cal D}$ is defined as a covariant derivative containing both gauge
fields,  ${\cal D}_\mu\qf_i=\prt_\mu\qf_i-A_\mu-a_\mu$.
The remarkable thing about (\ref{multform}) is that the action for the
edge bosons is completely contained in {\em bulk}
integrals. Derivatives of the step functions with respect to the $y$
coordinate lead to the edge terms in (\ref{Cbosonform}).
The constraint multiplied by $a_-$ is given by
\be
	0=\curl\left[\vec a-2p\;\sgn(m)\qy(y)\sum_i\vec{\cal D}\qf_i
	\right]
\label{constra}
\ee
with general solution
\be
	\vec{a}=\frac{2p}{2pm\qy(y)+1}\sgn(m)
	\left[\qy(y)\sum_i\vec{D}\qf_i+\nabla\qO\right]
\label{solconstr}
\ee
where $\qO$ is an arbitrary function.
Substitution of (\ref{solconstr}) into what remains of 
(\ref{multform}) after integration of $a_-$ yields an
expression
containing only the following linear combinations
\be 
	\tilde{\qf}_i \! := \! \qf_i \! -\! 2p\; \sgn(m)\qO.
\label{defft1}
\ee
The mapped action has a form similar to (\ref{multform}),
\bea
	S &=& \fr{i}{4\pi}\sum_{ij}
	\left[\qd_{ij}\;\sgn(m)-2p\fr{\nu}{m}\right]
	\int_\infty\left(2D_-\tilde{\qf}_i\curl
	[\qy(y)\vec{D}\tilde{\qf}_j]
	-\qy(y)\vec{D}\tilde{\qf}_i\!\times\!\prt_-
	\vec{D}\tilde{\qf}_j\right) \nn\\ &&
	-\fr{1}{2}(\fr{\nu}{2\pi m})^2\sum_{ij}
	\int_\infty U(\vec x-\vec x')\;
	\curl[\qy(y)\vec{D}\tilde{\qf}_i(\vec{x})]\;
	\nabla'\!\times\![\qy(y')\vec{D}\tilde{\qf}_j(\vec{x}')].
\label{mappedch1}
\eea
See appendix A for the explicit calculation. In the expression 
$[\cdots]$ we recognize the inverse of the coupling matrix 
(``$K$-matrix'')
that was obtained in Chern-Simons
theories,\cite{FrohlichZee,FrohlichKerler,FrohlichStuder,BlokWen1990,WenZee}
\be
	K_{ij}=\qd_{ij}\; \sgn(m)+2p.
\label{defKij}
\ee
Bringing the action (\ref{mappedch1}) into the form (\ref{Cbosonform})
we obtain
\bea
	S &=&
	\fr{i\nu}{4\pi}\int\!\! A\wedge dA
	-\fr{i}{4\pi}\sum_{ij}(K\inv)_{ij}\oint\left(
	D_x\tilde{\qf}_i D_-\tilde{\qf}_j -\tilde{\qf}_i E_x
	\right)\nn\\ &&
	-\fr{1}{2}(\fr{\nu}{2\pi m})^2\sum_{ij}
	\int_\infty U(\vec x-\vec x') \;
	\curl[\qy(y)\vec{D}\tilde{\qf}_i(\vec{x})]\;
	\nabla'\!\times\![\qy(y')\vec{D}\tilde{\qf}_j(\vec{x}')].
\label{mappedch2}
\eea
Eq. (\ref{mappedch2}) is the main result of this chapter. 
Even though bits and pieces of this action, such as the coupling
of $\tilde\qf$ to the gauge field\cite{Balachandran}, 
have appeared in the
literature before and the $K$-matrix structure is wholly familiar,
our complete theory is new, as many aspects of the impurity problem
were previously not understood.

The Hall conductance on the edge can unambiguously be determined from
(\ref{mappedch2}) by using the chiral
anomaly.\cite{Balachandran,Maeda}
The chiral anomaly enables us to use Laughlin's flux argument on the
edge, since it relates
the divergence of the edge current
to the electric field parallel to the edge,
\bea
	\prt_\mu J^\mu_{\rm edge}=-\fr{1}{2\pi}\qs_{\rm H}E_x
	& \hskip5mm;\hskip5mm & 
	J^\mu_{\rm edge}=\left.\fr{\qd S}{\qd A_\mu}\right|_{\rm edge}.
\eea
In this way we easily obtain that 
the Hall conductance on the edge
$\qs_{\rm H}$ is equal to $\nu$, as
it should be. 

In order to make contact with phenomenological models\cite{Wen}, we
mention that we
can write down a bulk Chern-Simons action that is completely equivalent to
(\ref{mappedch2}).
The electronic degrees of freedom are contained in fields $g_i$ that
act as potentials for the current,
\be
	S=\fr{i}{4\pi}\sum_{ij=1}^{|m|}
	(K\inv)_{ij}\left[-\int\! g^i\wedge dg^j
	+2\int\! g^i\wedge dA\eff\right]
	+\fr{\nu^2}{8\pi^2}\int\! U(\vec x-\vec x')B(\vec x) B(\vec x').
\label{potform}
\ee
The $g^i$ satisfy the following gauge fixing constraint at the edge 
\be
	\left[g_-^i(k_x)-i\fr{m}{\sqrt{2\pi}}U(k_x)\sum_{a=1}^{|m|}
	g_x^a(k_x)\right]_{\rm edge}=0.
\ee
The free part of (\ref{potform}) is
known from the work of Wen\cite{Wen}. 
However, lacking a microscopic edge theory, the full details of the
interactions were unknown.

\subsubsection{Charged and neutral modes}

Diagonalization of the $K$-matrix yields one charged mode and
$|m|-1$ neutral modes. The charged mode $\qG$ is given by
\be
	\qG=\fr{1}{|m|}\sum_i\tilde{\qf}_i
\label{charged}
\ee
where the normalization has been chosen such that $\qG$ has unit charge.
A possible choice of basis for the neutral modes $g_a$ is
\bea
\label{basisneutral}
	g_a&=& \fr{1}{a}\sum_{k=1}^a
	\tilde\qf_k-\tilde\qf_{a+1}
	\;\;\;\;\;\;\;\; a=1,\cdots,|m|-1
	\\
	\tilde\qf_{k}&=&\qG-\fr{k-1}{k}g_{k-1}+\sum_{a=k}^{|m|-1}
	\fr{1}{a+1}g_a. \nn
\eea
The Jacobian of the transformation from the $\tilde\qf_i$ to the
modes (\ref{charged}, \ref{basisneutral}) is unity.
In diagonalized form the action reads
\bea
	S &=& \fr{i\nu}{4\pi}\left[
	\int\!\! A\wedge dA
	-\oint\left(D_x\qG D_-\qG-\qG E_x\right)
	\right] 
	-\fr{i}{4\pi}\sgn(m)\sum_{a=1}^{|m|-1}\fr{a}{a+1}
	\oint \prt_x g_a\prt_- g_a
	\nn\\&&
	-\fr{1}{2}(\fr{\nu}{2\pi})^2
	\int_\infty U(\vec x-\vec x')\;
	\curl[\qy(y)\vec{D}\qG(\vec{x})]\;
	\nabla'\!\times\![\qy(y')\vec{D}\qG(\vec{x}')].
\label{Sdiag}
\eea
Naturally, the charged mode is the only one coupling to $A_\mu$
and feeling interactions.

\subsubsection{(In)stability of counter-flowing channels}
\label{secinstability}

Here we make an important observation which has been missed in the
literature, namely that {\em for the noninteracting case or for short-range
interactions the Chern-Simons
mapping is ill-defined at negative} $m$. 
In order to see this, we look at the Hamiltonian for the charged mode 
corresponding to (\ref{Sdiag})
in the absence of Coulomb interactions,
\be
	H_{\qG}^{\rm (free)}
	=\frac{\vd\nu}{4\pi}\oint\! dx\; (D_x\qG)^2.
\label{Hamilg}
\ee
Recall that  $\vd$ is defined as $m/(2\pi\qr_{\rm edge})$
and that $\nu\! >\! 0$. Hence,
for negative $m$ the free Hamiltonian is unbounded from below!
The Coulomb interactions cure this situation by
adding an extra term to the edge Hamiltonian,
\be
	H_{\rm c}=\fr{1}{2}(\fr{\nu}{2\pi})^2\oint \! dxdx'\;
	U(x,x')D_x\qG(x)D_x\qG(x')
\label{Hamilc}
\ee
effectively changing the velocity $\vd$ to 
\bea
	\tilde{v}(k_x)=\vd+\fr{\nu}{\sqrt{2\pi}}U(k_x)
	& \hskip5mm,\hskip5mm &
	U(k_x)\propto \ln(k_x)
\label{defvtilde}
\eea
For low momenta the logarithm in $\tilde{v}(k_x)$ is dominant,
ensuring that the effective velocity $\tilde{v}$ is positive even if
$\vd\!<\!0$ and thereby
yielding a bounded Hamiltonian.
We conclude that the long range Coulomb interactions are actually
needed in order for the Chern-Simons mapping to be well defined. A
similar situation occurs in the bulk theory with half integer filling
fraction [I] where in the absence of long range interactions the CS
procedure gives rise to a divergent quasiparticle d.o.s.

\ns{Tunneling density of states}
\label{secshorttunn}
\subsection{The integer case}
In [III] we obtained the following results for the IQHE.
In the $Q$-field theory, the tunneling d.o.s. for tunneling electrons from a 
Fermi liquid into the edge of a quantum Hall system is given by
\be
	\langle Q^{\qa\qa}(\qt_2,\qt_1,\vec{x}_0) \rangle :=
	\sum_{n=-\infty}^{\infty}e^{i\nu_n(\qt_2-\qt_1)}
	\langle Q_{nn}^{\qa\qa}(\vec{x}_0)\rangle
\label{Qtunnel}
\ee
where $\vec x_0$ is the position where the tunneling electron enters the 
sample, $\qt_1$ and $\qt_2$ are two instants in imaginary time and
$\nu_n$ is the bosonic Matsubara frequency $\fr{2\pi}{\qb}n$.
The process of integrating out $Q$ from the action (\ref{Qform})
changes (\ref{Qtunnel}) to 
\be
	\left\langle\exp -i\left[\prt_c\inv A_c\eff(\qt_2,\vec{x}_0)
	-\prt_c\inv A_c\eff(\qt_1,\vec{x}_0)\right]\right\rangle.
\label{Atunnel}
\ee
The process of decoupling the quadratic edge term in $A_\mu$
by the introducing of boson fields $\qf_i$ transforms (\ref{Atunnel}) into 
\be
	\left\langle\exp -i\int_{\qt_1}^{\qt_2}\! d\qt\;\prt_0
	\qf_j(\qt,\vec x_0)\right\rangle,
	\;\;\;\;\; j=1,\ldots,|m|
\label{expphi}
\ee
where the expectation value is now taken with respect to the $\qf$ action
(\ref{Cbosonform}). 
Any of the fields $\qf_1,\cdots,\qf_{|m|}$ can be put in the exponent,
since they are interchangeable.

The completely equivalent edge theories in terms of $Q$ (\ref{Qform}) and in
terms of $\qf_i$ (\ref{Cbosonform}) now ensure that the expressions
(\ref{Qtunnel}) and (\ref{expphi}) are identical. An explicit
computation leads to the same Fermi liquid result
\bea
	(\qt_2-\qt_1)^{-S} &\hskip5mm ;\hskip5mm & S=1.
\eea
in both cases.

\subsection{The fractional case}
\label{sectunndosfrac}
\subsubsection{Results of the Chern-Simons mapping}

We next wish to extend the CS mapping procedure to include the result
for the electron propagator at the edge (\ref{expphi}). 
This problem is nontrivial since the amount of statistical flux that
is trapped now depends on whether the path from
time $\qt_1$ to $\qt_2$ is taken just outside or just inside the
sample. Hence, the obvious translation of (\ref{expphi}) into
composite fermion language,
\be
	\left\langle \exp -i\int_{\qt_1}^{\qt_2}\!\!\! 
	d\qt\; [\prt_0\qf_j-a_0](\qt,\vec x_0) \right\rangle,
\label{add_a0}
\ee
now has an additional {\em geometrical} significance.
In order to find out what this geometrical significance is, we must
repeat the steps (\ref{multform}--\ref{mappedch2}), but now in the
presence of (\ref{add_a0}).

We write $a_0=a_- +iv a_x$. The presence of
$a_-$ at $\vec x_0$ modifies the constraint equation 
(\ref{constra}) and we get
\be
	\curl\left[\vec a+2p\;\sgn(m)\qy(y)\sum_i(\vec a-\vec D\qf_i)\right]
	=-2p\cdot2\pi\qd(\vec x-\vec x_0)L
\label{modifiedconstr}
\ee
where we have defined a quantity $L$ such that
\be
	\qb\sum_n L_{-n} f_n(\vec x)=\int_{\qt_1}^{\qt_2}\!\!
	d\qt f(\qt,\vec x)
\label{defL}
\ee
for any bosonic function $f$.
The factor of $2\pi$ in (\ref{modifiedconstr}) originates from our
convention $\hbar\!=\!1$, while a flux quantum is given by $h/e$,
and the minus sign shows that the statistical flux indeed points in
the negative $z$ direction.
From (\ref{modifiedconstr}) we see that the amount of flux is given by
\be
	2p\;\; {\rm for }\;\; y_0<0 \hskip5mm ; \hskip5mm
	2p\nu/m \;\; {\rm for }\;\; y_0>0.
\label{chargenu/m}
\ee
This is an important result. It shows that for $y_0 \!>\! 0$
{\em the CS mapping has altered the charge of the tunneling particle} 
from~1 into $\nu/m$. A physical interpretation is readily given in
terms of Laughlin's gauge argument. If one flux quantum is sent
through a hole in the IQH sample, one electron will leave the edge per
filled Landau level, resulting in a total charge $m$. The CS
mapping changes the transported charge, which is equal to the Hall conductance,
to $\nu$. Since this charge is contained in $m$ Landau levels, the
charge per Landau level has to be $\nu/m$.

We will now integrate out the Chern-Simons gauge field.
The general solution to the constraint equation
(\ref{modifiedconstr}) is 
\be
	\vec a(\vec x)=
	\fr{2p\;\sgn(m)}{2pm\qy(y)+1}\left[\qy(y)\sum_i\vec D\qf_i
	+\nabla\qO+\sgn(m) L \nabla\arctg\fr{x-x_0}{y-y_0}
	\right]
\ee
with $\qO$ again an arbitrary gauge.
Substituting this solution $\vec a$ into what remains of the action, 
we re-obtain the action (\ref{mappedch2}), but now with a slightly
different definition of the $\tilde\qf$ fields,
\be
	\tilde\qf_i=\qf_i-2p\;\sgn(m)\qO-2pL\; \arctg\fr{x-x_0}{y-y_0}.
\label{deftildephi}
\ee
See appendix~B for the details of this calculation.
The expectation value acquires the following form after the CS mapping
\be
	\exp  -i\int_{\qt_1}^{\qt_2}\!\!\! 
	d\qt\! \left\{
	\prt_0\tilde\qf_i-\qy(y_0)2p\fr{\nu}{|m|}
	\sum_j D_0\tilde\qf_j \right\}(\qt,\vec x_0).
\label{expwiththeta}
\ee
For further considerations we put $A_\mu\!=\!0$, since the tunneling is done at
zero gauge field.

Several remarks need to be made here. First of all, we see that
the outcome will indeed not depend on the gauge $\qO$.
Secondly, the expectation value will be different for
$\vec x_0$ going to the edge from inside or
from outside of the sample. Furthermore, there is a difference between the
cases $m\!>\!0$ and $m\!<\!0$.
 
If we take $y_0\! \uparrow \!0$, we easily obtain the result 
(see appendix C)
\be
	\left\langle \exp -i \int_{\qt_1}^{\qt_2}\!\!\! d\qt\; \prt_0
	\tilde\qf_i \right\rangle = (\qt_2-\qt_1)^{-S_{\rm out}}
	\hskip5mm ; \hskip5mm
	S_{\rm out} = 2p+1+\left(\fr{1}{m}-\fr{1}{|m|}\right)
\label{Soutside}
\ee
which is the tunneling exponent found by \cite{KFP}.

For positive $m$ we have $S_{\rm out}\!=\!K_{ii}\!=$odd, 
which clearly indicates a composite fermion.
For negative $m$ the meaning of the exponent is not clear; it does not
in general represent a fermionic quantity.

For $y_0\! \downarrow \!0$ we can rewrite (\ref{expwiththeta}) as
\be
	\exp \left[ -i\;\sgn(m) \int_{\qt_1}^{\qt_2}\!\!\! 
	d\qt\; \prt_0\sum_j K\inv_{ij}\tilde\qf_j(\qt,\vec x_0)
	\right]
\label{expinside}
\ee
Expression (\ref{expinside}) gives us a tunneled charge equal to
$\sgn(m)\sum_j K_{ij}\inv=\nu/m$
in accordance with (\ref{chargenu/m}).
Note that the charge is not contained in one Landau channel, as was the
case in the tunneling expression for the integer effect, but
instead spread out over all the channels. In the $i$'th channel there
is a charge $1-2p\nu/|m|$, in all the other channels $-2p\nu/|m|$.

It is readily seen that the tunneling exponent becomes
\be
	S_{\rm in}=1-2p\fr{\nu}{m}+\left(\fr{1}{m}-\fr{1}{|m|}\right).
\label{Sinside}
\ee
For positive $m$ this expression takes the form 
$S_{\rm in}\!=\!K\inv_{ii}\!=\! 1\!-\! 2p\nu/m$.

\subsubsection{Duality}

Notice that  the cases $y_0\! \uparrow \!0$ and
$y_0\! \downarrow \!0$ are related by a ``T-duality'' in the sense of
ref.\cite{Balachandran}, with $K$ playing the role of compactification
radius. 
If we define ``dual'' fields $\xi_i$ with charge $\nu/m$ as
$\xi_i\!=\!\sgn(m)\; K_{ij}\inv\tilde\qf_j$, then the
action (\ref{mappedch2}) takes the form 
\bea
	S[\xi]&=&\fr{i\nu}{4\pi}\int\! A\wedge dA-\fr{i}{4\pi}\sum_{ij}
	K_{ij}\oint \left(D_x\xi_i D_-\xi_j-\fr{\nu}{m}\xi_i E_x\right)
	\nn\\ &&
	-\fr{1}{2}(\fr{1}{2\pi})^2\sum_{ij}
	\int_\infty U(\vec x-\vec x') \;
	\curl[\qy(y)\vec{D}\xi_i(\vec{x})]\;
	\nabla'\!\times\![\qy(y')\vec{D}\xi_j(\vec{x}')],
\label{Sxi}
\eea
with $D_\mu\xi_i \!=\! \prt_\mu\xi_i \!-\! \fr{\nu}{m}A_\mu$.
The exponent (\ref{expinside}) is written as 
\be
	\exp [-i\int_{\qt_1}^{\qt_2}\!\! d\qt\; \prt_0\xi_i],
\label{expdual}
\ee 
i.e. the same form as
(\ref{Soutside}) but now in terms of $\xi$.
The action (\ref{Sxi}) for the $\xi$ fields contains the matrix $K$
instead of $K\inv$. 
In \cite{Balachandran} it was shown that the energy spectrum for a
system of 1+1D coupled bosons is
invariant under O$(m,m,Z)$ transformations acting on a specific 
$2m \!\times\! 2m$ matrix that expresses the Hamiltonian in terms of
the momenta and winding numbers of a state.
The replacement $K\!\naar\! K\inv$ is an example of such an O$(m,m,Z)$
transformation. In a certain way it interchanges fluxes and particles.

In the above considerations the statistics parameter is calculated of a
particle with charge one; the amount of statistical flux that it
acquires is either $2p$ (outside the sample) or $2p\nu/m$ (inside). 
If we were, on the other hand, to
concentrate on the flux instead of the charge, we would consider the
case of $2p\nu/m$ statistical flux quanta which acquire {\em charge},
namely $1$ inside the sample and $\nu/m$ outside. 
In this sense there is charge-flux duality.

\begin{figure}
\begin{center}
\setlength{\unitlength}{1mm}
\begin{picture}(80,35)(0,0)
\put(0,5)
{\epsfxsize=80mm{\epsffile{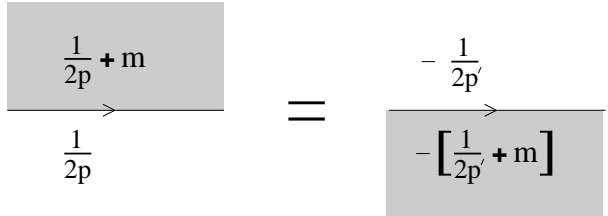}}}
\end{picture}
\caption{Inverse number of statistical flux quanta inside and outside
the sample. The shaded area denotes the inside.
The duality $p \!\leftrightarrow\! p'$ reverses the role of the
opposite sides of the edge.}
\label{figpprime}
\end{center}
\end{figure}

Notice that the roles of inside/outside get reversed in the argument
given above. We make this more explicit by  defining
\be
	p'=-p\fr{\nu}{m}.
\ee
The duality relation $K\!\naar\! K\inv$ can be written as
\be
	K(p',m)=K\inv(p,m).
\ee
In Fig.~\ref{figpprime} it is shown that an exchange of $p$ to $p'$
interchanges the opposite sides of the edge.
We will come back to this point in section \ref{secdual}.


In the case of the simple Laughlin fractions $\nu\!=\!1/(2p\!+\!1)$
the duality is particularly transparent. The fact that there is only
one edge channel means that there is no redistribution of charge over
various channels, which allows for an easy interpretation of the
tunneling exponent in terms of flux-charge composites. 
Let us consider the complex conjugate of (\ref{expinside}); taking the
complex conjugate does not change the tunneling exponent, but it
changes a particle to a hole. 
Expression (\ref{expinside}) becomes 
$\exp -i\int \!d\qt(-\nu)\prt_0\tilde\qf$, 
describing the tunneling of a hole of charge $-\nu$.
The tunneling exponent is given by $S_{\rm in}\!=\!\nu$. We can
interpret this result as follows. The hole has a statistical flux
$-2p\nu$ attached to it. In addition, it has an ``intrinsic'' flux of
$-\nu$ coming from the fermionic statistics of the electron. The
statistical parameter of a charge-flux composite is given by the
product of charge and flux, in this case 
$(-\nu)\cdot(-2p\nu\!-\!\nu)\!=\!\nu$
for the quasihole.
The relation $K\!\naar\! K\inv$ takes the simple form 
$\nu\inv\!\naar\!\nu$.


\ns{Long range disorder}
\label{seclongrange}
\subsection{Separation of edge channels; Edge particles}
\label{secseparation}

In [III] we discussed how
long range disorder can cause the edge states of different Landau
levels to become spatially separated. A potential fluctuation at the edge
generically lifts all states in such a way that new `edges' are created.
If the chemical potential lies between the
shifted and unshifted energy of a Landau level, the edge states of this
Landau level will be situated inside the sample, not on the outermost edge.
If there are several potential jumps of this kind, all the edge channels
can become separated.
Viewed from above, the edge channels will mostly lie closely together at
the edge, while at certain points one or more of them go
venturing into the bulk.

This picture leads us to the idea of describing the chiral bosons by {\em
one} field $\qf(\vec x)$ that lives on $|m|$ edges instead of $|m|$ fields
that live on one edge.
The action for the integer effect (\ref{Cbosonform}) then becomes
\bea
	S&=&\fr{i}{4\pi}\left[\int \! nA^\qa\wedge dA^\qa
	-\sgn(m)\sum_{a=1}^{|m|}\oint_a\!(D_x\qf D_-\qf
	-E_x\qf) \right] \nn\\
	&& -\fr{1/2}{(2\pi)^2}\int_{x,x'}U(\vec x-\vec x')
	\curl[n(\vec x)\vec{D}\qf(\vec x)]\nabla'\!\times\!
	[n(\vec x')\vec{D}\qf(\vec x')]
\label{localfillings}
\eea
where $n(\vec x)$ stands for the local filling: outside the sample
$n(\vec x)$  
is zero; every time you cross an edge it changes its value by one, until it
reaches its bulk value $m$.
The $\oint_a$ denotes integration over the $a$'th edge.
Every edge is given a label $a\in Z$, equal to the maximum filling
that the edge is bordering on.
For positive $m$ we have $a\!\in\! \{1,\cdots,m\}$, for negative $m$
$a\!\in\! \{m\!+\!1,\cdots,0\}$.

\begin{figure}
\begin{center}
\setlength{\unitlength}{1mm}
\begin{picture}(100,100)(0,0)
\put(0,5)
{\epsfxsize=100mm{\epsffile{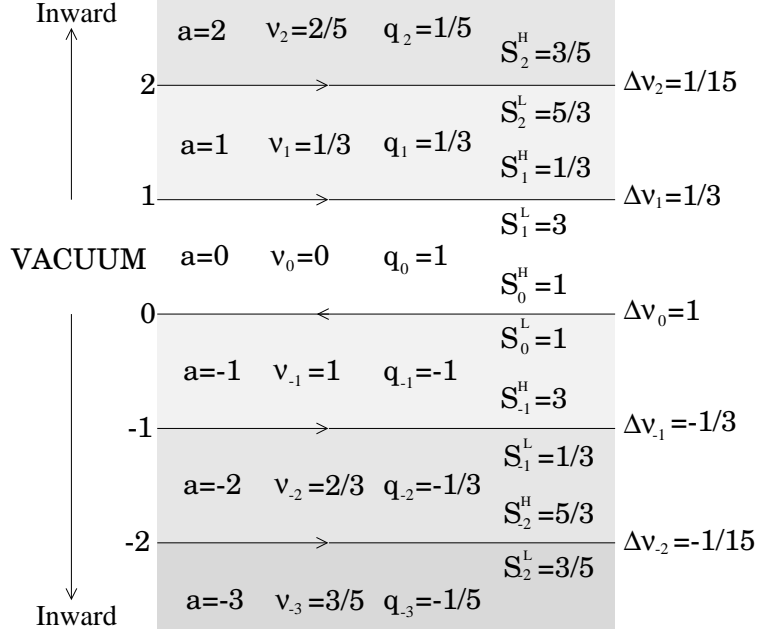}}}
\end{picture}
\caption{Edge quasiparticle duality for $2p=2$. 
A schematic drawing is given of the regions with different 
$\nu(\vec x)$; systems with $m\!<\!0$ and
$m\!>\!0$ are combined in one diagram. Arrows represent the chirality
of an edge. Given are the local filling fraction, the charge of the
tunneling particle as well as the
tunneling exponent $S$ on both sides of each edge.}
\label{figalledges}
\end{center}
\end{figure}

So we now have regions in which the integer effect occurs at different
values, separated by edges of the $\nu=1$ type on which the boson field
$\qf$ lives. Performing the Chern-Simons mapping to the fractional effect
simply maps the filling fractions in these regions separately from
$n(\vec x)$ to $\nu(\vec x)$,
\be
	\nu(\vec x)=\fr{n(\vec x)}{2pn(\vec x)+1}.
\ee
The mapped action reads
\bea
	S&=&\fr{i}{4\pi}\left[\int \! \nu A\wedge dA
	-\sum_{a=1}^{|m|}\triangle\nu_{[a]}
	\oint_a\!(D_x\qF D_-\qF
	-E_x\qF) \right] \nn\\
	&& -\fr{1/2}{(2\pi)^2}\int_{x,x'}U(\vec x-\vec x')
	\curl[\nu(\vec x)\vec{D}\qF(\vec x)]\nabla'\!\times\!
	[\nu(\vec x')\vec{D}\qF(\vec x')].
\label{Sseparated}
\eea
[A redefinition of the type (\ref{deftildephi}) has occurred and the
new field is denoted as $\qF$.]
Integration over the $a$'th edge counting from the vacuum 
is written as $\oint_a$.
We use the label $a$ both for edges and for regions: the region lying
``above'' the $a$'th edge also carries the label $a$.
The $\dn_{[a]}$ is defined as $\nu_a\!-\!\nu_{a-1}$ for positive $m$
and $\nu_{-a}\!-\!\nu_{-(a-1)}$ for negative $m$; 
here $\nu_a$ stands for
the local filling fraction in region number $a$.
In Fig.~\ref{figalledges}  we have made an overview 
(for samples with both signs of $m$ in the same figure)
of the parameters $a$, $\nu_a$ and
$\triangle\nu_a$ in the simple case $p=1$.  

The tunneling term is simply a generalization of (\ref{expwiththeta}),
\be
	\left\langle \exp -i \int_{\qt_1}^{\qt_2}\! d\qt\;
	[1-2p\nu(\vec x_0)] \prt_0\qF(\qt,\vec x_0)\right\rangle,
\label{isolatedexp}
\ee
where the expectation value is taken
with respect to (\ref{Sseparated}). The result again depends on the
direction from which the edge is approached. 
The tunneled charge is given by $q_a\!=\! \nu_a/a$ if an edge is
approached from region number $a$. 

Evaluating (\ref{isolatedexp}) is  a nontrivial exercise, since
the Coulomb interaction couples the edges to each other.

If we imagine the edges to be far apart, only the intra-edge
interaction is still present, drastically simplifying the calculation.
Tunneling exponents are then given by the simple expression
$S\!=\!q^2/\triangle\nu$.
Labeling an approach from above by `H' and from below by `L', we write
the tunneling exponents at the $a$'th edge ($a\!\neq\! 0$) as follows
\bea
	S^{\rm L}_a=1+2p\fr{\nu_{a-1}}{a-1}=
	\fr{a}{\nu_a}\cdot\fr{\nu_{a-1}}{a-1} &  & \nn\\
	S^{\rm H}_a=1-2p\fr{\nu_a}{a}
	=\fr{\nu_a}{a}\cdot\fr{a-1}{\nu_{a-1}} 
	& \hskip5mm ; \hskip5mm &
	S^{\rm L}_a S^{\rm H}_a=1.
\label{isolatedS}
\eea
The edge $a\!=\!0$ represents the outer boundary of a negative-$m$
sample. Here we get
\be
	S^{\rm H}_0=S^{\rm L}_0=2p-1.
\ee
(See Fig.~\ref{figalledges}.)
Notice that in this approach, with spatially separated edge channels,
the vacuum
exponent for $m\!<\!0$ is always fermionic. 
Notice also that the channel labeled $a\!=\!0$, lying between the
vacuum and an $m\!<\!0$ sample, is the only one that needs long-range
Coulomb interactions in order to be stable. In Fig.~\ref{figalledges}
this channel has a chirality opposite to all the others.

\subsection{Enlarged duality}
\label{secdual}
\begin{figure}
\begin{center}
\setlength{\unitlength}{1mm}
\begin{picture}(80,55)(0,0)
\put(0,5)
{\epsfxsize=80mm{\epsffile{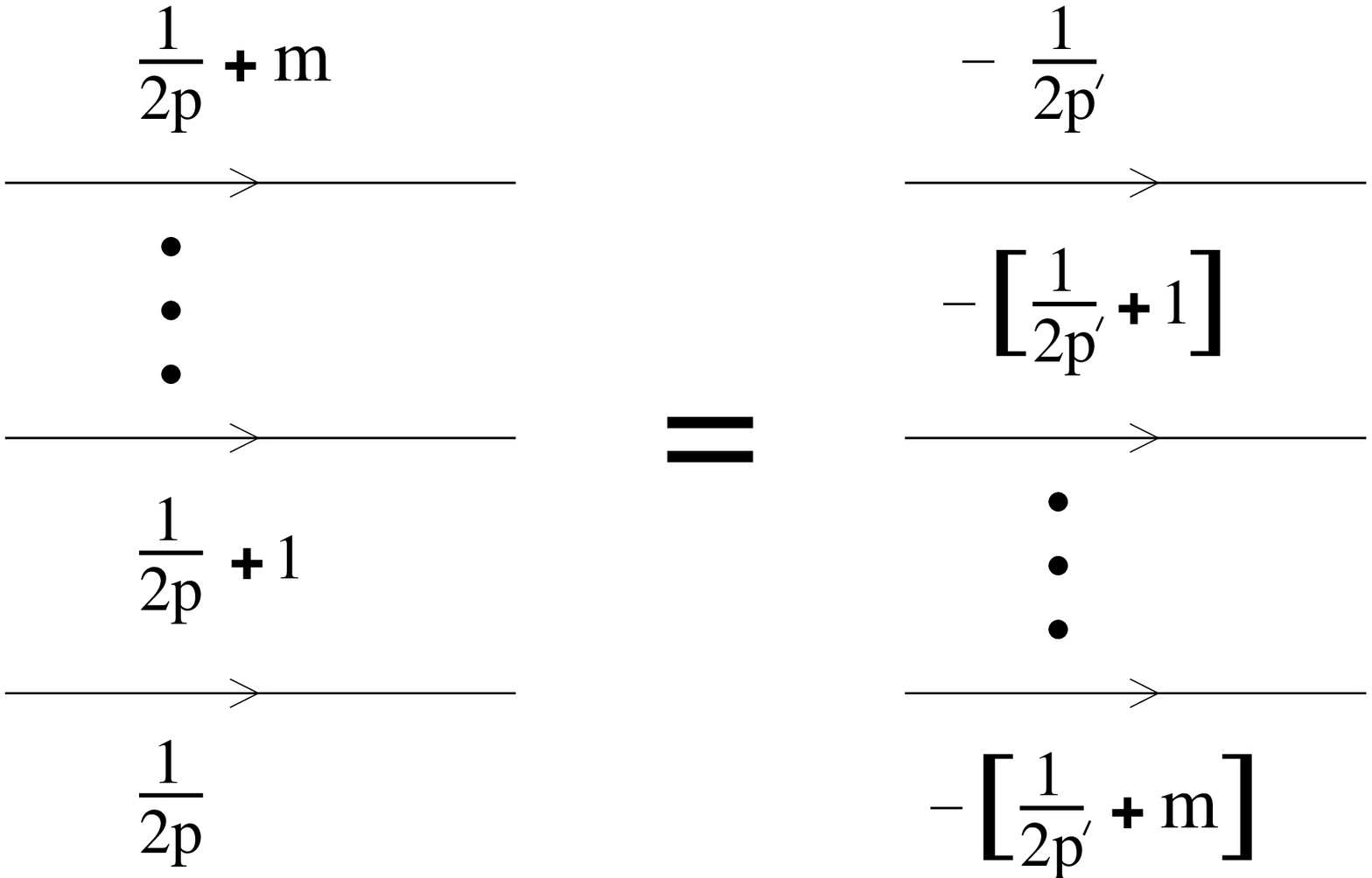}}}
\end{picture}
\caption{Geometrical interpretation of the duality relation
$K\!\!\naar\! K\inv$ as a reversal of edge channels.
For each region the inverse number of statistical flux quanta
is given. ($m\! > \! 0$)}
\label{figDuality}
\end{center}
\end{figure}

As mentioned in section~\ref{sectunndosfrac}, 
the Hamiltonian of the chiral boson
model has a duality symmetry under $K\!\naar\! K\inv$.
In the case of spatially separated edge channels, this duality
transformation has an extra geometrical interpretation:
{\em The order of the edge channels gets reversed}.
The transformation $K\!\naar\! K\inv$ can be accomplished by transforming
$p\!\naar\! p'$, with
\be
	2p'=-2p\nu /m.
\label{defp'}
\ee
The easiest way to visualize the reversal of channels is to consider
how the transformation (\ref{defp'}) affects the amount of statistical
flux that a tunneling particle carries (see Fig.~\ref{figDuality}). 
The inverse number of flux quanta is $1/2p$ outside the sample and
increases with unit steps every time an edge is crossed, until the
value $m\!+\!1/2p$ is reached in the bulk proper. Writing (\ref{defp'}) as
$-\frac{1}{2p'}\!=\!\frac{1}{2p}\!+\!m$, we see that 
\be
	-(\fr{1}{2p'}+a)=\fr{1}{2p}+(m-a).
\ee
The minus sign preserves the chirality of the edge particles.

More generally, under a transformation 
\be
	p\naar p'=-p\nu_b/b,\;\;\;\;\;\; b\in Z
\ee
the whole structure of Fig.~\ref{figalledges}, including region
labels, edge labels, filling fractions and tunneling exponents, gets
mirrored in the line $a\!=\!a_{\rm mirror}$,
\be
	a_{\rm mirror}=\frac{b+1}{2}.
\ee
This results in the following relations between tunneling exponents at
different edges:
\bea
\label{mirroredS}
	S^{\rm H}_a(p')=S^{\rm L}_{b+1-a}(p) & \hskip5mm ; \hskip5mm&
	S^{\rm L}_a(p')=S^{\rm H}_{b+1-a}(p) \\
	a\neq 0 & \hskip5mm ; \hskip5mm & b+1-a \neq 0. \nn
\eea
Unless $b\!=\!-1$, expression (\ref{mirroredS}) does not work for
the mirrored tunneling exponent of the
counter-flowing edge $a=0$ and its mirror image $a\!=\!b\!+\!1$.
\be
	S^{\rm H}_{b+1}(p')=S^{\rm L}_{b+1}(p')=1-2p=-S_0(p)
\ee 
\be
	S_0(p')=-(1+2p\nu_b/b) =  \left\{\matrix{
	 -S^{\rm L}_{b+1}(p) &
	\mbox{for }b\neq -1 \cr
	 S_0(p) & \mbox{for }b=-1}\right. \nn
\ee
There are two special cases: 
\begin{itemize}
\item
$b\!=\!-1$, where the symmetry axis lies exactly on the counter-flowing
channel. Every region gets mapped into minus itself, while the flux is
left unchanged! ($p' \!=\!p$.) This shows that at given $p$, all tunneling
exponents are symmetric under ($a\!\naar\! -a$, L$\leftrightarrow$H).
The vacuum gets mapped into the $a\!=\!-1$ region, which is suggestive of
an interpretation of this region as a filled 
$\nu\!=\!1/(2p\!-\!1)$ `vacuum' from
which holes can be created.
\item
$b\!=\!0$, with the symmetry axis at $a\!=\!1/2$. 
We have $p' \!=\! -p$, so we are
in a situation with reversed magnetic field. This transformation
relates samples with composite fermion filling $m$ to those with $-m$. 
\end{itemize}

\subsection{Relation between dirty and clean edges}
\label{secPhiphi}
We considered the various edge modes to be far apart. The results for
the exponents (\ref{isolatedS}) are valid provided that one probes the
system at length scales short relative to the distance between the
edge modes. In this section, we consider the action (\ref{Sseparated})
in the opposite limit of large distances. The edge modes then strongly
interact and (\ref{isolatedS}) now becomes a Luttinger liquid theory
of a sharp, ``clean'' edge. It is not a priori obvious that this
theory is physically the same as the theory of the sharp ``dirty''
edge. 
In the following we show that this is indeed the case.
In particular, the two theories yield identical results 
for the two kinds of
tunneling exponent that can be found from (\ref{expwiththeta}) in the
``dirty'' $\tilde\qf$-theory, namely the case where the limit to the edge is
taken from the vacuum and the case where one approaches the edge from
inside the sample.

If one writes $\qF_i$ for the field $\qF$ evaluated on the $i$'th edge
then it is possible to express the fields $\qF_i$ in terms of a charged
mode $\qG$ and neutral modes $\qg_1,\cdots,\qg_{|m|-1}$ in such a way
that the action (\ref{Sseparated}) for the $\qF_i$ exactly matches 
the action (\ref{mappedch2}) for the $\tilde\qf_i$. (In principle, the
charged and neutral modes are not locally defined fields, since they
contain parts from different edges. However, in the long distance
limit all the edges are compressed together on one contour and the
fields are local.)
The relation between the $\qF_i$ and the $\tilde\qf_i$ is nontrivial;
we show that the fields are dual to each other for $m\!>\!0$
and explain why this leads to identical numbers when one calculates
tunneling exponents according to the clean picture 
(\ref{isolatedexp}) (see Fig.~\ref{figalledges}) or the dirty
scenario (\ref{expwiththeta}); we also show
that there is some extra structure for negative $m$ which is the
source of the differing 
exponents as obtained from 
(\ref{expwiththeta}) versus (\ref{isolatedexp}) at $m\!<\!0$.

\subsubsection{Example for positive $m$: $\nu=2/5$}
\begin{figure}
\begin{center}
\setlength{\unitlength}{1mm}
\begin{picture}(130,50)(0,0)
\put(0,5)
{\epsfxsize=130mm{\epsffile{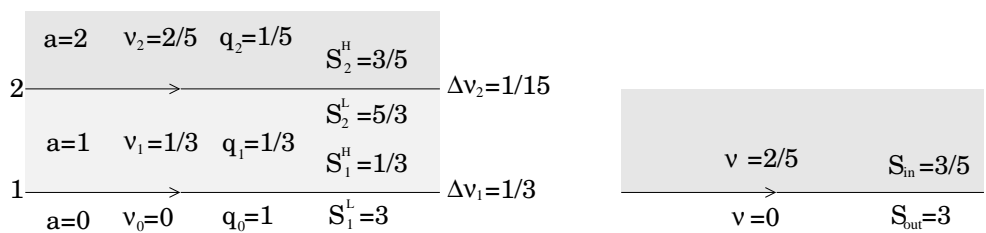}}}
\end{picture}
\caption{Clean and dirty scenario for the 
$\nu\!=\!2/5$ edge. Left: spatially separated counter-flowing channels.
Right: result of the $K$-matrix theory.
Listed are filling fractions, charges and tunneling exponents.}
\label{figedge25}
\end{center}
\end{figure}

First we give a simple example for $\nu\!=\!2/5$ ($m\!=\!2$,
$p\!=\!1$). In both $\tilde\qf$ and $\qF$-theory the tunneling
exponents for this filling are given by $S_{\rm out}\!=\!3$, 
$S_{\rm in}\!=\! 3/5$ (see Fig.~\ref{figedge25}).

For $\nu\!=\!2/5$ the $\tilde\qf$-action is given by
(\ref{mappedch2}), with the $K$-matrix as defined in
(\ref{defKij}). Written explicitly we have for $A_\mu\!=\!0$
\be
	S_K=-\fr{i}{4\pi}\sum_{ij}\left(
	\matrix{\fr{3}{5} & -\fr{2}{5} \cr
		-\fr{2}{5} & \fr{3}{5}}\right)_{ij}
	\oint\! \prt_x\tilde\qf_i\prt_-\tilde\qf_j
	-\half (\fr{1/5}{2\pi})^2\sum_{ij}\oint_{xx'}\!\!
	U(x-x')\; \prt_x\tilde\qf_i(x)\prt_{x'}\tilde\qf_j(x').
\ee
The diagonalized form (\ref{Sdiag}) of this action is given by
\be
	S_K=-\fr{i}{4\pi}\oint\left[\fr{2}{5}\prt_x\qG\prt_-\qG
	+\half\prt_x g\prt_- g \right]-\half(\fr{2/5}{2\pi})^2
	\oint_{xx'}\!\!
	U(x-x')\; \prt_x\qG(x)\prt_{x'}\qG(x')
\label{SK25}
\ee
where the charged and neutral mode are defined in (\ref{basisneutral})
and in this case read
\bea
	\left(\matrix{g \cr \qG}\right)=
	\left(\matrix{1 & -1 \cr \half & \half}\right)
	\left(\matrix{\tilde\qf_1 \cr \tilde\qf_2}\right)
	& \hskip5mm ; \hskip5mm & 
	\left(\matrix{\tilde\qf_1 \cr \tilde\qf_2}\right)=
	\left(\matrix{\half & 1 \cr -\half & 1}\right)
	\left(\matrix{g \cr \qG}\right).
\label{ffromg25}
\eea
On the other hand, the action for the $\qF$-field on the separated
edges is given by (\ref{Sseparated}). 
If we compress all the edges together we get (again for $A_\mu\!=\!0$)
\bea
	S_{\rm sep} &=& -\fr{i}{4\pi}\oint\left[
	\fr{1}{3} \prt_x\qF_1\prt_-\qF_1
	+\fr{1}{15} \prt_x\qF_2\prt_-\qF_2\right] \nn\\
	&& -\half(\fr{1}{2\pi})^2\oint_{xx'}\! U(x-x')
	\prt_x(\fr{1}{3}\qF_1+\fr{1}{15}\qF_2)
	\prt_{x'}(\fr{1}{3}\qF_1+\fr{1}{15}\qF_2).
\label{SF25}
\eea
We define a neutral mode $\qg$ and charged mode
$\qG$ for the $\qF$-theory as follows
\bea
	\left(\matrix{\qg \cr \qG}\right)=
	\left(\matrix{1 & -1 \cr \fr{5}{6} & \fr{1}{6} }\right)
	\left(\matrix{\qF_1 \cr \qF_2}\right)
	& \hskip5mm ; \hskip5mm & 
	\left(\matrix{\qF_1 \cr \qF_2}\right)=
	\left(\matrix{\fr{1}{6} & 1 \cr -\fr{5}{6} & 1}\right)
	\left(\matrix{\qg \cr \qG}\right).
\label{FfromG25}
\eea
Whereas the $\tilde\qf$ fields are all equivalent in the dirty
edge theory, 
here the $\qF$ are not, since the spatially separated edges have
unequal $\dn$ values which appear in the definition of $\qG$.
In terms of $\qG$ and $\qg$ the $\qF$-action 
(\ref{SF25}) takes the form
\be
	S_{\rm sep}=-\fr{i}{4\pi}\oint\left[
	\fr{2}{5}\prt_x\qG\prt_-\qG+\fr{1}{18}\prt_x\qg\prt_-\qg
	\right]
	-\half(\fr{2/5}{2\pi})^2
	\oint_{xx'}\!\!
	U(x-x')\; \prt_x\qG\prt_{x'}\qG
\label{Ssep25}
\ee
Comparing the two results (\ref{SK25}) and (\ref{Ssep25}), we see that
there is complete equivalence given the relation
\be
	g=\fr{1}{3}\qg.
\label{gamma=3g}
\ee
Having shown that the actions are equivalent,
we now want to compare tunneling operators. 
In order to do this, we in principle only need to rewrite the 
$\tilde\qf$ and $\qF$ that appear in the various operators in terms of
charged and neutral modes, and then use 
(\ref{SK25}) or (\ref{Ssep25}). However, much insight can be gained by
also expressing $\tilde\qf$ and $\qF$ in terms of each other.
Combining (\ref{ffromg25}), (\ref{FfromG25}) and (\ref{gamma=3g}), we find
\bea
	\left(\matrix{\tilde\qf_1 \cr \tilde\qf_2}\right)=
	\left(\matrix{ 1 & 0 \cr \fr{2}{3} & \fr{1}{3} }\right)
	\left(\matrix{\qF_1 \cr \qF_2}\right)
	& \hskip5mm ; \hskip5mm & 
	\left(\matrix{\qF_1 \cr \qF_2}\right)=
	\left(\matrix{ 1 & 0 \cr -2 & 3}\right)
	\left(\matrix{\tilde\qf_1 \cr \tilde\qf_2}\right).
\label{ffromF}
\eea
In the iqHe $\qf$-theory we can choose which one of the fields $\qf_1$
or $\qf_2$ to put in the tunneling operator. Eq.~(\ref{expwiththeta}) 
then gives four 
possible operators, which we list in the table below,
giving only the 
``$X$'' in the expression $\exp -i X|_{\qt_1}^{\qt_2}$.

\begin{center}
\begin{tabular}{| l | c | c |}
\hline
& $\tilde\qf$-basis & Diag. basis \\ \hline
Vacuum & $\qf_1\naar \tilde\qf_1$ & 
	$\hphantom{\fr{1}{5}} \qG+\fr{1}{6}\qg$ \\ \cline{2-3}
       & $\qf_2\naar \tilde\qf_2$ & 
	$\hphantom{\fr{1}{5}} \qG-\fr{1}{6}\qg$ \\ \hline
Bulk   & $\qf_1\naar(K\inv\tilde\qf)_1
=\hphantom{-}\fr{3}{5}\tilde\qf_1-\fr{2}{5}\tilde\qf_2$ &
$\fr{1}{5}\qG+\fr{1}{6}\qg$ \\ \cline{2-3}
       & $\qf_2\naar(K\inv\tilde\qf)_2
=-\fr{2}{5}\tilde\qf_1+\fr{3}{5}\tilde\qf_2$ &
$\fr{1}{5}\qG-\fr{1}{6}\qg$ \\ \hline
\end{tabular}
\end{center}

\vskip5mm

There are only two corresponding operators in the $\qF$-theory. 
An approach from the vacuum can only involve $\qF_1$ because the other
field does not live on the outermost edge. In the same way, the bulk
operator can only contain $\qF_2$. 
From (\ref{isolatedexp}) we get

\begin{center}
\begin{tabular}{| l | c | c | c |}
\hline
& $\qF$-basis & $\tilde\qf$-basis & Diag. basis \\ \hline
Vacuum & $\hphantom{\fr{1}{5}} \qF_1$ & 
	$\tilde\qf_1$ & $\hphantom{\fr{1}{5}} \qG+\fr{1}{6}\qg$ 
 \\ \hline
Bulk   & $\fr{1}{5}\qF_2$ & 
$-\fr{2}{5}\tilde\qf_1+\fr{3}{5}\tilde\qf_2$ & 
$\fr{1}{5}\qG-\fr{1}{6}\qg$ \\ \hline
\end{tabular}
\end{center}

\vskip5mm

In all cases the exponents obtained from the operators in these tables
are $S_{\rm out}\!=\! 3$, $S_{\rm in}\!=\! 3/5$.
There is no a priori reason why these numbers should be the same as
those appearing in Fig.\ref{figedge25}. In the short distance theory
the actions for the two edges are uncoupled, while we have now used
the action (\ref{SK25}) where they are coupled.

\subsubsection{Generalization for $m\!>\!0$}
The analysis for $\nu\!=\! 2/5$ is easily generalized to all filling
fractions resulting from positive $m$. Let us redo all the steps in
the same order.

The $K$-matrix action for the $\tilde\qf_i$ is given by
(\ref{mappedch2}). Its diagonal form is given by (\ref{Sdiag}) with
$\sgn(m)\!=\!+1$ and $\qG$, $g_1,\cdots,g_{m-1}$ as defined in
(\ref{basisneutral}).
For the $\qF$-theory we define the charged and neutral modes as
follows
\bea
	&& \qG=\fr{1}{\nu}\sum_{a=1}^{m}\dn_a\qF_a
	 \hskip5mm ; \hskip5mm 
	\qg_k=\fr{1}{\nu_k}\sum_{a=1}^k \dn_a\qF_a-\qF_{k+1}
	\nn \\ 
	&& \qF_a=\qG+\sum_{k=a}^{m-1}\fr{\dn_{k+1}}{\nu_{k+1}}\qg_k
	-\fr{\nu_{a+1}}{\nu_a}\qg_{a-1}.
\label{defdiag}
\eea
The diagonalized form of the action reads
\be
	S= -\fr{i}{4\pi}
	\oint\left[ \nu\prt_x\qG\prt_-\qG
	+\sum_{k=1}^{m-1}\fr{k}{k+1}q_k^2
	\prt_x\qg_k\prt_-\qg_k \right]
	-\fr{\nu^2}{8\pi^2}\oint_{xx'} U\; \prt_x\qG\prt_{x'}\qG
\label{Sdiag2}
\ee
The only difference with (\ref{Sdiag}) is the factor $q_k^2$ in front
of the neutral modes. The actions for $\tilde\qf$ and $\qF$ are
identical if we define
\be
	g_k=q_k\qg_k.
\label{g=qgamma}
\ee
Using (\ref{basisneutral}), (\ref{defdiag}) and (\ref{g=qgamma}) we
are now able to relate the $\qF_i$ to the $\tilde\qf_i$,
\be
	q_k\qF_k=\sum_{a=1}^k(K_{k\times k}\inv)_{ka}\tilde\qf_a.
\label{qF=Kinvf}
\ee
This is a remarkable result. It shows that the fields are dually
related. Notice that the $K$-matrix appearing in (\ref{qF=Kinvf}) does
not have the full $m \!\times\! m$ size. In order to find the $k$'th
$\qF$-field  one has to use the $K$-matrix for a truncated system
which has only $k$ edges, i.e. 
$(K_{k\times k}\inv)_{ij}\!=\! \qd_{ij} \!-\! 2p q_k$.
If one wants to write (\ref{qF=Kinvf}) as a vector equation of the
form
$(q\qF)_i \!=\! M_{ij}\tilde\qf_j$, the $M$ is {\em not} given by
$K\inv$ but by a stacking together of rows taken from matrices 
$K_{k\times k}\inv$ of increasing size.

Given below are those tunneling operators for the clean
and dirty scenario 
that describe the same tunneling processes. 

\vskip5mm
\begin{center}
\begin{tabular}{| l | c | c |} \hline
& $\tilde\qf$-theory & $\qF$-theory \\ \hline
Vacuum & $\qf_k\naar \tilde\qf_k$ & $\qF_1=\tilde\qf_1$ \\ \hline
Bulk & $\qf_k\naar (K\inv\tilde\qf)_k$ & 
	$q_m\qF_m= (K\inv\tilde\qf)_m$ \\ \hline
\end{tabular}
\end{center}

From appendix~C we know that the $\tilde\qf_1,\cdots,\tilde\qf_m$ all
give the same result $S_{\rm out}\!=\!2p\!+\!1$. 
That settles it for the vacuum operators. The
beautiful thing about (\ref{qF=Kinvf}) is that it ensures that the
bulk operator $q_m\qF_m$ takes the form of the truly dual field 
$(K\inv\tilde\qf)_m$ involving the full $K$-matrix. From
(\ref{Sinside}) we know that all the dual fields yield the same
exponent $S_{\rm in}\!=\! 1\!-\!2pq_m$.

So now we have shown for $m\!>\!0$ that the long distance limit of the
short distance $\qF$-theory indeed reproduces the physics of the 
$\tilde\qf$-theory.
But why does the $\qF$-theory at positive $m$
also give the same exponents {\em before the long distance limit is
taken}? The answer is again provided by (\ref{qF=Kinvf}).
Eq.~(\ref{isolatedS}) shows that the exponent on the isolated $m$'th edge
is equal to  $q_m^2/\dn_m\!=\! q_m/q_{m-1}$. 
On the other hand, the long distance picture leads to operators 
$\exp -i(K\inv\tilde\qf)_i$  in terms of fields dual to $\tilde\qf_i$ which
have coupling matrix $K$ instead of $K\inv$. This leads to an exponent
$K_{ii}\inv$, which is exactly equal to $q_m/q_{m-1}$.

\subsubsection{Example for negative $m$: $\nu=2/3$}
\begin{figure}
\begin{center}
\setlength{\unitlength}{1mm}
\begin{picture}(130,50)(0,0)
\put(0,5)
{\epsfxsize=130mm{\epsffile{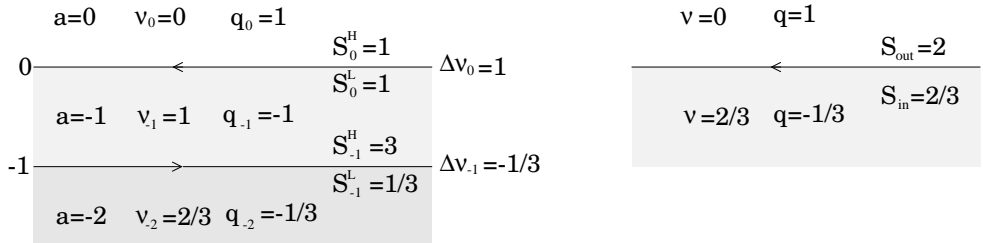}}}
\end{picture}
\caption{Clean and dirty scenario for the
$\nu\!=\!2/3$ edge. Left: spatially separated counter-flowing channels.
Right: $K$-matrix theory.
Listed are filling fractions, charges and tunneling exponents.}
\label{figedge23}
\end{center}
\end{figure}

For $m\!<\!0$ the relation between the clean and dirty scenario
is more subtle. In order to get the flavor, let us first
consider the easiest example, $\nu\!=\!2/3$. In Fig.\ref{figedge23} we
have shown the relevant piece of Fig.\ref{figalledges} as compared to
the $K$-matrix picture.

Our starting point for the $\tilde\qf$-theory is again
(\ref{mappedch2}). The explicit action for $A_\mu\!=\!0$ is given by
\be
	S_K= -\fr{i}{4\pi}\sum_{ij}
	\left(\matrix{-\fr{1}{3} & \fr{2}{3} \cr 
	\fr{2}{3} & -\fr{1}{3}}\right)_{ij}
	\oint\! \prt_x\tilde\qf_i\prt_-\tilde\qf_j
	-\half(\fr{1/3}{2\pi})^2\sum_{ij}\oint_{x,x'}U
	\prt_x\tilde\qf_i\prt_{x'}\tilde\qf_j
\ee
and after diagonalization it takes the form
\be
	-\fr{i}{4\pi}\oint\left[\fr{2}{3}\prt_x\qG\prt_-\qG
	-\half\prt_x g\prt_- g\right]
	-\half(\fr{2/3}{2\pi})^2\oint_{x,x'}U\prt_x\qG\prt_{x'}\qG.
\label{SK23}
\ee
The definition of the charged and neutral modes again follows from
(\ref{basisneutral}),
\bea
	\left(\matrix{g \cr \qG}\right)=
	\left(\matrix{1 & -1 \cr \half & \half}\right)
	\left(\matrix{\tilde\qf_1 \cr \tilde\qf_2}\right)
	& \hskip5mm ; \hskip5mm & 
	\left(\matrix{\tilde\qf_1 \cr \tilde\qf_2}\right)=
	\left(\matrix{\half & 1 \cr -\half & 1}\right)
	\left(\matrix{g \cr \qG}\right).
\label{ffromg23}
\eea
The action (\ref{Sseparated}) for the $\qF$-theory with compressed
edges is now given by
\be
	S_{\rm sep}=-\fr{i}{4\pi}\oint\left[\prt_x\qF_1\prt_-\qF_1
	-\fr{1}{3} \prt_x\qF_2\prt_-\qF_2\right] 
	-\fr{1/2}{(2\pi)^2}\oint_{x,x'}U 
	\prt_x(\qF_1-\fr{1}{3}\qF_2)\prt_{x'}(\qF_1-\fr{1}{3}\qF_2)
\ee
Charged and neutral modes are defined in a way
similar to (\ref{defdiag}), taking into account the different signs of
the $\dn$.
\bea
	\left(\matrix{\qg \cr \qG}\right)=
	\left(\matrix{1 & -1 \cr \fr{3}{2} & -\fr{1}{2} }\right)
	\left(\matrix{\qF_1 \cr \qF_2}\right)
	& \hskip5mm ; \hskip5mm & 
	\left(\matrix{\qF_1 \cr \qF_2}\right)=
	\left(\matrix{-\fr{1}{2} & 1 \cr -\fr{3}{2} & 1}\right)
	\left(\matrix{\qg \cr \qG}\right).
\label{FfromG23}
\eea
This leads to the following diagonalized form,
\be
	S_{\rm sep}=
	-\fr{i}{4\pi}\oint\left[
	\fr{2}{3}\prt_x\qG\prt_-\qG-\half\prt_x\qg\prt_-\qg
	\right]-\half(\fr{2/3}{2\pi})^2\oint_{x,x'}U\prt_x\qG\prt_{x'}\qG
\label{Ssep23}
\ee
We find that (\ref{SK23}) and (\ref{Ssep23}) are identical after defining
$g=\qg$. Using (\ref{ffromg23}) and (\ref{FfromG23}) we get
\bea
	\left(\matrix{\tilde\qf_1 \cr \tilde\qf_2}\right)=
	\left(\matrix{ 2 & -1 \cr 1 & 0 }\right)
	\left(\matrix{\qF_1 \cr \qF_2}\right)
	& \hskip5mm ; \hskip5mm & 
	\left(\matrix{\qF_1 \cr \qF_2}\right)=
	\left(\matrix{ 0 & 1 \cr -1 & 2}\right)
	\left(\matrix{\tilde\qf_1 \cr \tilde\qf_2}\right).
\label{Ffromf23}
\eea
This relation does {\em not} satisfy (\ref{qF=Kinvf}) for the
quantities $q_{-1}\qF_1$ and $q_{-2}\qF_2$. (Negative indices have to
be taken because the relevant regions in Fig.\ref{figedge23} are
$a\!=\!-1$ and $a\!=\!-2$.)
Instead we have
\bea
	q_{-1} \qF_1 &=& \tilde\qf_1 -2\qG
	\nn\\
	q_{-2} \qF_2 &=& (K\inv\tilde\qf)_2 -\fr{2}{3}\qG.
\eea
Eq. (\ref{qF=Kinvf}) produces the first term on the r.h.s.,
an expression that has the right
magnitude of the charge but the wrong sign. The $q_{-1} \qF_1$ has charge
$-1$, so the extra term $-2\qG$ is precisely what is needed to flip
the sign of the charge. In the same way, the term $-\fr{2}{3}\qG$
changes the charge from the $+1/3$ that $(K\inv\tilde\qf)_2$ yields to
the $-1/3$ that is correct for the $q_{-2}\qF_2$.

An interesting subtlety is that even though (\ref{qF=Kinvf}) does not
hold for the relation between $\qF$ and $\tilde\qf$ fields, all
tunneling exponents behave as if it did hold. 
The reason for this is the following. 
After a tunneling operator has been expressed in terms of charged and
neutral modes, it is easy to see that the only effect of
the extra term involving $\qG$ is to change the sign of the factor
standing in front of $\qG$. This sign change has no effect whatsoever
on the calculation of the tunneling exponent.
In the table below we explicitly write down all the tunneling
operators.

\begin{center}
\begin{tabular}{| l | c | c || c | c | c |}
\hline
& \multicolumn{2}{c||}{$\tilde\qf$-theory} &
\multicolumn{3}{c|}{$\qF$-theory} \\ \hline
& $\tilde\qf$-basis & Diag. basis &
$\qF$-basis & $\tilde\qf$-basis & Diag. basis \\ \hline
Vacuum & $\qf_1\naar\tilde\qf_1$ & $\qG+\half\qg$ &
$\qF_1$ & $\tilde\qf_2$ & $\qG-\half\qg$ \\ \cline{2-6}
& $\qf_2\naar\tilde\qf_2$ & $\qG-\half\qg$ &
& & \\ \hline
Bulk & $\qf_1\naar\fr{1}{3}\tilde\qf_1-\fr{2}{3}\tilde\qf_2$ & 
$-\fr{1}{3}\qG +\half\qg$ &
$-\fr{1}{3}\qF_2$ & $\fr{1}{3}\tilde\qf_1-\fr{2}{3}\tilde\qf_2$ & 
$-\fr{1}{3}\qG+\half\qg$ \\ \cline{2-6}
& $\qf_2\naar -\fr{2}{3}\tilde\qf_1+\fr{1}{3}\tilde\qf_2$ & 
$-\fr{1}{3}\qG-\half\qg$ &
& & \\ \hline
\end{tabular}
\end{center}

\vskip5mm

In both the $\qF$-theory and $\tilde\qf$-theory the tunneling
exponents obtained from the above expressions are
\bea
	S_{\rm out}=2 &\hskip5mm ; \hskip5mm &
	S_{\rm in}=2/3.
\eea
The discrepancy with the short distance clean scenario result 
$S_{\rm out}\!=\!1$, $S_{\rm in}\!=\! 1/3$ is caused by the fact that 
(\ref{Ffromf23}) does not satisfy (\ref{qF=Kinvf}), so that the
argument given in the positive $m$ subsection does not apply.

\subsubsection{Generalization for arbitrary sign of $m$}
For completeness, we now present general equations that comprise all
the above cases.
The $\tilde\qf$-theory expressions (\ref{Sdiag}) and
(\ref{basisneutral}) are equally valid for negative and positive $m$. No
extra work is needed there.
The $\qF$-theory expressions (\ref{defdiag}) and (\ref{Sdiag2}) do
need some modification if we want to incorporate the possibility of
$m$ being negative.
Eq. (\ref{defdiag}) for the charged and neutral modes can be generalized as
follows,
\bea
	&& \qG=\fr{1}{\nu}\sum_{a=1}^{|m|}\dn_{[a]}\qF_a
	 \hskip5mm ; \hskip5mm 
	\qg_k=\fr{1}{\nu_{[k]}}\sum_{a=1}^k \dn_{[a]}\qF_a-\qF_{k+1}
	\nn \\ 
	&& \qF_a=\qG+\sum_{k=a}^{|m|-1}\fr{\dn_{[k+1]}}{\nu_{[k+1]}}\qg_k
	-\fr{\nu_{[a+1]}}{\nu_{[a]}}\qg_{a-1}
\eea
Here the notation $\nu_{[a]}$ (with $a\!>\!0$) means 
$\nu_a$ for samples with positive $m$ and $\nu_{-a}$ for $m\!<\!0$
samples. For instance, in case of negative $m$ we would have
$\dn_{[2]}\!=\!\nu_{-2}\!-\!\nu_{-1}$.

Using this definition, the generalization of (\ref{Sdiag2}) becomes
\be
	S= -\fr{i}{4\pi}
	\oint\left[ \nu\prt_x\qG\prt_-\qG
	+\sgn(m)\sum_{k=1}^{|m|-1}\fr{k}{k+1}q_{[k]}^2
	\prt_x\qg_k\prt_-\qg_k \right]
	-\fr{\nu^2}{8\pi^2}\oint_{xx'} U\; \prt_x\qG\prt_{x'}\qG
\ee
Comparing this expression to (\ref{Sdiag}) we can make the
identification 
\be
	g_k=|q_{[k]}| \qg_k.
\ee
This enables us to relate the fields $\tilde\qf_i$ to the $\qF_a$.
\bea
	\tilde\qf_k &=& 
	\sgn(m)\left[q_{[k]}\qF_k+\sum_{a=1}^{k}2p\dn_{[a]}\qF_a\right]
	+[1-\sgn(m)]\cdot\qG
	\\
	q_{[k]}\qF_k &=& \sgn(m)\tilde\qf_k
	-2pq_{[k]}\sum_{a=1}^{k}\tilde\qf_a +[1-\sgn(m)]\cdot q_{[k]}\qG
	\nn\\ &=&
	\sum_{a=1}^k (K_{k\times k}\inv)_{ka}\tilde\qf_a
	+[1-\sgn(m)]\cdot q_{[k]}\qG
\label{generalFf}
\eea
where $K_{k\times k}$ again denotes the $K$-matrix for a system containing
$k$ edges. It is easily checked that for positive $m$ (\ref{qF=Kinvf})
is reobtained. 

Eq. (\ref{generalFf}) shows that what happened at $\nu\!=\!2/3$
happens at all negative $m$ states: The $K\inv\tilde\qf$ part of the
tunneling expression $q_m\qF_{|m|}$
has the right magnitude of the charge $q_m$ but the
wrong (=positive) sign. The extra term $2q_m\qG$ flips the
sign. The reason why we write $\qG$ in (\ref{generalFf}) instead of
expressing it in terms of the $\tilde\qf_a$ is that we can now directly discern
the effect on the tunneling operators. In the diagonal basis the
effect of the extra $\qG$ term is to flip the sign of the coefficient
of $\qG$ in the tunneling operators. Just as in the $2/3$ case this
leaves the exponent unaffected.

Also generalized to all $m\!<\!0$,
the discrepancy between the exponents as calculated in the
short and long distance limits is due to the fact that
(\ref{qF=Kinvf}) does not hold,

\subsection{Quasiparticles and the Laughlin argument}
\label{quasiLaughlin}
The particles probed in tunneling experiments are not
necessarily the same as those featuring in Laughlin's flux
argument. 

In the integer case,
the effect of putting a flux quantum through a
hole in the sample is a transport of one electron through each Landau
level below the Fermi energy. In the language of
(\ref{localfillings}),
considering for simplicity $m\!>\!0$,
what would happen at the $a$'th edge is that $a\!-\!1$ electrons enter
that edge from below and that $a$ electrons leave it on the upper
side. In this way, a total of $m$ electrons gets transported into the
bulk, which is the correct number. The operator describing this
process is given by
\be
	\exp -i\int_{\qt_1}^{\qt_2}\!\! d\qt\left\{
	\vphantom{\int}
	a[\prt_0\qf-a_0](\vec x_{\rm H})
	-[a-1][\prt_0\qf-a_0](\vec x_{\rm L})\right\}
\ee
where $\vec x_{\rm H}$ lies infinitesimally above the $a$'th edge 
and $\vec x_{\rm L}$ lies below it. Both points have the same $x$-coordinate.

Knowing the results of section~\ref{secseparation}, it is easy to do the
Chern-Simons mapping. The constraint equation multiplied by $a_-$
contains the sum of two delta functions instead of just
one. Correspondingly, the redefined field $\qF$ contains the sum of
two $\arctg$ functions.
The mapped expression is given by
\be
	\exp -i\int_{\qt_1}^{\qt_2}\!\! d\qt\left\{
	\vphantom{\int}
	\nu_a \prt_0\qF(\vec x_{\rm H})
	-\nu_{a-1}\prt_0\qF(\vec x_{\rm L})\right\}.
\ee
As is clearly seen  from the constants appearing here,
the emitted charge is $\nu_a$ and the absorbed charge is $\nu_{a-1}$.
Taking all the edges together we have a total transported charge of
$\sum_{a=1}^m \triangle\nu_a=\nu_m$, which is the correct number for a
system of filling fraction $\nu_m$.
Taking the limit where $\vec x_{\rm H}$ and $\vec x_{\rm L}$ go to the
same point $x$ on the edge, the tunneling expression takes the simple
form
\be
	\exp -i\int_{\qt_1}^{\qt_2}\!\! d\qt\; 
	\triangle\nu_a\prt_0\qF(\qt,x).
\ee

\ns{Edge tunneling at arbitrary $\nu$}
\label{seclongtunn}
In [III] we discussed how the 
non-quantized tunneling exponent\cite{continuum} $1/\nu$ can be
understood as a consequence of the Coulomb interactions between the
edge modes and the localized states present in the bulk. 
For $\nu$ around integer filling, we explicitly showed how
the interactions render the neutral modes irrelevant. 
Here we will generalize this result to fractional fillings, in particular to
Jain's main hierarchy.

At almost integer filling $\nu\!=\!m\!+\!\qe$, a fraction $\qe$ of the total
area is occupied by regions with filling $m\!+\!1$, while the rest of the
sample has filling $m$. After performing the flux attachment, a
fraction $1\!-\!\qe$ of the area has filling $\nu_m$ and a fraction $\qe$
has $\nu_{m+1}$, leading to
\be
	\nu=m+\qe \hskip3mm
	\stackrel{\mbox{CS mapping}}{\longrightarrow}\hskip3mm 
	\nu=\nu_m+\qe(\nu_{m+1}-\nu_m).
\ee

\begin{figure}
\begin{center}
\setlength{\unitlength}{1mm}
\begin{picture}(150,60)(0,0)
\put(0,5)
{\epsfxsize=150mm{\epsffile{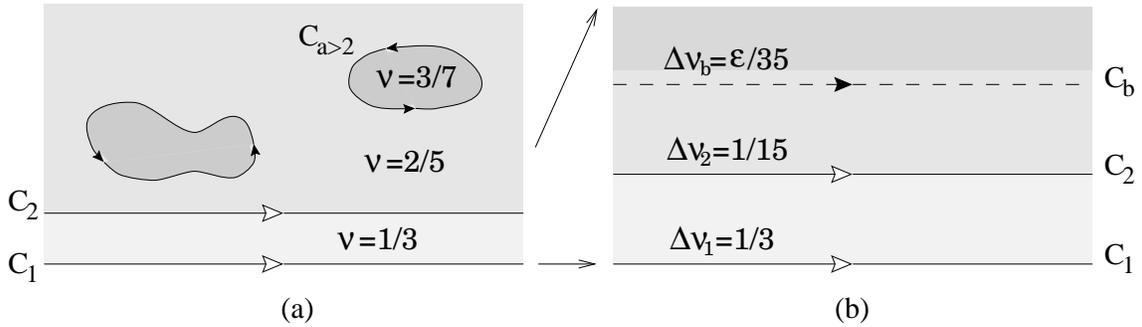}}}
\end{picture}
\caption{Example of the long wavelength limit
for filling fraction $\nu\!=\!2/5 \!+\!\qe/35$. 
(a) The $3/7$ islands
occupy a fraction $\qe$ of the total area.
(b) Long wavelength limit. The bulk degrees of freedom are effectively
represented
by an extra chiral boson on the contour $C_b$ with charge density 
$\fr{\qe}{35}\prt_x\qF|_{C_b}$. The filling fraction is expressed as
$\nu\!=\!\triangle\nu_1 \!+\! \triangle\nu_2 \!+\! \triangle\nu_b$.}
\label{figIslands}
\end{center}
\end{figure}

The action for this situation is given by (\ref{Sseparated}), but now
including the bulk contours $C_{a>m}$.
In [III] we showed that in the long wavelength limit the
bulk degrees of 
freedom can effectively be described by an extra chiral boson on a
contour $C_b$ near the edge plus an edge term
\be
	S_{\rm n}=
	-\qa\int_{C_b}dk\; |k| \prt_\perp\qF(-k)\prt_\perp\qF(k)
\label{Schop}
\ee
(with $\prt_\perp$ the derivative normal to $C_b$ and $\qa$ a positive
constant), 
which suppresses the neutral modes.
The only difference between the integer and fractional case lies in
the edge parameters $\triangle\nu_1,\cdots,\triangle\nu_m$
 which were all equal to 1 in the integer case and
$\triangle\nu_b \! :=\!\qe(\nu_{m+1}-\nu_m)$ which was equal to $\qe$.
In terms of fields $\qF_1,\ldots,\qF_m$ and 
$\qF_{m+1} \! :=\!\qF_b$ which live on {\em one} edge, the long wavelength
limit yields the following action
\bea
	S[\qF] &=& -\fr{i}{4\pi}\sum_{a=1}^{m+1} \qd_a\oint\prt_x\qF_a
	\prt_-\qF_a - \fr{1}{8\pi^2}\sum_{a,b=1}^{m+1}\qd_a\qd_b
	\oint_{xx'}\!\! 
	U(x-x')\; \prt_x\qF_a\prt_{x'}\qF_b
	\nn\\
	&& -\fr{1}{8\pi^2}\sum_{a,b=1}^{m+1} V_{ab}\qd_a\qd_b\oint
	\prt_x \qF_a\prt_x\qF_b +S_{\rm n}.
\label{Slimit}
\eea
Here we have defined 
$\qd_a\!=\!\triangle\nu_a$ for $a\!=\!1,\ldots,m$
and $\qd_{m+1}\!=\!\qe\triangle\nu_{m+1}$.
All short range effects, originating from the fact that the bulk
orbitals sometimes venture near the edge,
are contained in the matrix $V$.

The definition of a charged mode
$\qG$ and neutral modes $\qg_1,\ldots, \qg_m$ is similar to
(\ref{defdiag}),
\bea
	\qG  =  \fr{1}{\nu}\sum_{a=1}^{m+1}\qd_a\qF_a 
	& \hskip5mm ; \hskip5mm &
	\qg_k=\fr{1}{\nu_k}\sum_{a=1}^k \qd_a\qF_a -\qf_{k+1}.
\eea
The $\qF$'s are expressed in terms of the $\qg$'s as follows
\be
	\qF_a = \qG-\frac{\nu_{a-1}}{\nu_a}\qg_{a-1}
	+\sum_{k=a}^m \frac{\qd_{k+1}}{\nu_{k+1}}\qg_k.
\ee
Written in the new basis, the action (\ref{Slimit}) is given by
\bea
	S[\qG,\qg] &=& S_{\rm n}[\qg]
	-\frac{i}{4\pi}\oint\left[
	\nu\prt_x\qG\prt_-\qG+\sum_{k=1}^m\qd_{k+1}
	\frac{\nu_k}{\nu_{k+1}}\prt_x\qg_k\prt_0\qg_k
	\right] \nn\\
	&& -\frac{\nu^2}{8\pi^2}
	\oint_{xx'}\!\! U(x,x')\prt_x\qG
	\prt_{x'}\qG
	-\frac{1}{8\pi^2}\sum_{a,b=1}^{m}\oint\! 
	V_{ab}\;\prt_x\qg_a\prt_x\qg_b.
\eea
The neutral modes do not contribute to a calculation of the tunneling
density of states 
because of (\ref{Schop})
and consequently the parameter $\nu$ in front of the
$\prt_x\qG\prt_-\qG$ term determines the tunneling exponent.
The result is $S=1/\nu$, the same result that we obtained for the
integer theory.


\ns{Plateau transitions}
\label{secPlattrans}
We have discussed integer plateau transitions in [III].
Recapitulating the results obtained there, we can sketch
a transition from $\nu\!=\!m$ to $\nu\!=\!m\!+\!1$ (with $m$ positive) as
follows. One has $m$ copies of the $\nu\!=\!1$ type $Q$-edge theory on
the edge of the sample and one copy percolating through the bulk.
Upon taking the continuum limit, the percolating ``edge'' gives rise
to a nonlinear sigma model with $\qs_{xx}\!=\!\half$ and
$\qs_{xy}\!=\!\half$. Combining the edge and bulk contributions, one
gets a plateau transition at
$\qs_{xx}\!=\!\half$, $\qs_{xy}\!=\!m\!+\!\half$.

We now generalize this picture to the fractional case, in particular
to the transition $\nu_n\!\naar\!\nu_{n+1}$. 
At the boundary of the sample we have $n$ edge channels with total
contribution 
$(\nu_1\!-\!\nu_0)\!+\!(\nu_2\!-\!\nu_1)
\!+\!\cdots\!+\!(\nu_n\!-\!\nu_{n-1})\!=\!\nu_n$
to the Hall conductance.
One edge channel with parameter 
$\dn\!=\!\nu_{n+1}\!-\!\nu_n$ (see Fig.\ref{figIslands}a)
percolates through the bulk.
In order to find out what kind of sigma-model arises from this
percolating ``edge'', we 
first we go back to the $Q$-field formalism from (\ref{Sseparated}).
As mentioned in 
section~\ref{secbosint}, we obtained a theory for chiral bosons
starting from an action for the matrix field $Q$ coupled to the
external field $A_\mu$. 
For completeness, we give the the action $S_{\dn}[Q,A]$
for one edge channel with parameter $\triangle\nu$, 
\bea
	S_{\dn}[Q,A] &=& S_{\rm top}[Q]
	+\fr{i}{4\pi}\left[\sum_\qa\int\!
	\nu(A\eff)^\qa\wedge d(A\eff)^\qa-\dn\ointdx A_x\dagg A_c\eff
	\right] \nn\\
	&& +\fr{\pi}{4\qb\vd}S_{\rm F}[Q]-\fr{\pi}{4\qb}
	\sum_{n\qa}\intd{k_x}\fr{1}{\tilde v(k_x)}\left|\tr\Iamn Q
	-\fr{\qb}{\pi}\sqrt{\dn}(A_c\eff)^\qa_n\right|^2 \nn\\
	&& +\fr{\qb}{2}(\fr{1}{2\pi})^2\intd{^2 x d^2 x'}
	\nu(\vec x)B(\vec x)U(\vec x-\vec x')\nu(\vec x')B(\vec x')
\label{mappedQ}
\eea
where in the definition of $A_0\eff$ (\ref{defA0eff}) the $m$ now
has to be replaced by $\nu(\vec x)$, and in the definition of $\tilde v$ 
(\ref{defvtilde}) one should read $\dn$ instead of $\nu$.
We have neglected here the Coulomb interaction with other channels.

Notice that in (\ref{mappedQ}) the charge of the $Q$-field is no
longer 1 but $\sqrt{\dn}$. A gauge transformation takes the form
\bea
	A_\mu\naar A_\mu+\prt_\mu\chi &\hskip5mm ; \hskip5mm &
	Q\naar e^{i\sqrt{\dn}\cdot\hat\chi}Qe^{-i\sqrt{\dn}\cdot\hat\chi}.
\eea
Following [III] we use (\ref{mappedQ}) and obtain the $\qs$ model
action for the percolating network. Putting $A_\mu\!=\!0$, the result
for the kinetic terms is equal to
\be
	S_{\rm cont}[Q]= -\fr{1}{8}\tilde\qs_{xx}\Tr (\nabla Q)^2
	+\fr{1}{8}\tilde\qs_{xy}\qe^{ij}\Tr Q\prt_i Q\prt_j Q
\label{percsigmas}
\ee
where
\be
	\tilde\qs_{xx}=\half \hskip5mm ; \hskip5mm
	\tilde\qs_{xy}=\half
\ee
now represent the mean field conductances of the {\em composite
fermions}. The {\em actual} mean field values obtained from linear
response in the external fields are different by a factor of 
$\dn$, i.e.
\bea
	\qs_{xx}'=\half\dn \hskip5mm ;\hskip5mm
	\qs_{xy}^{\rm bulk}=\half\dn
\label{realcond}
\eea
Adding the edge contribution to this result, we finally get for the
mean field conductances at the 
plateau transition
\bea
	\qs_{xx}'=\half\cdot \fr{1}{(2pn+p+1)^2-p^2}
	&\hskip5mm ; \hskip5mm&
	\qs_{xy}'=\half(\nu_n+\nu_{n+1})=
	\fr{(2pn+p+1)(n+1/2)-p/2}{(2pn+p+1)^2-p^2}.
\label{ntransition}
\eea
This result differs from what one would expect from acting with Sl(2,Z)
on an integer plateau transition. In general, the Sl(2,Z) mapping of
the conductances takes the form 
\be
	\qs\naar\qs' = \fr{\qs}{2p\qs+1}, 
\label{SL2Z}
\ee
where the complex number $\qs$ is defined as 
$\qs\!=\!\qs_{xy}\!+\! i\qs_{xx}$.
In terms of $\qs_{xy}$ and $\qs_{xx}$ (\ref{SL2Z}) reads
\bea
	\qs_{xx}'=\fr{\qs_{xx}}{(2p\qs_{xy}+1)^2+(2p\qs_{xx})^2}
	&\hskip5mm ; \hskip5mm&
	\qs_{xy}'=\fr{\qs_{xy}(2p\qs_{xy}+1)+2p\qs_{xx}^2}
	{(2p\qs_{xy}+1)^2+(2p\qs_{xx})^2} 
\eea
Inserting the composite fermion values 
$\qs_{xx}\!=\!\half$, $\qs_{xy}\!=\! n\!+\!\half$ we now obtain
\bea
	\qs_{xx}'=\half\cdot\fr{1}{(2pn+p+1)^2+p^2}
	&\hskip5mm ; \hskip5mm &
	\qs_{xy}'= \fr{(2pn+p+1)(n+1/2)+p/2}{(2pn+p+1)^2+p^2}.
\label{fromSL2Z}
\eea
The different results (\ref{ntransition}) and (\ref{fromSL2Z})
indicate that the Chern-Simons mapping is not uniquely described by
Sl(2,Z) but depends on the details of disorder.


\ns{Summary and conclusions}
\label{secsummary}
We have derived, from first principles, the complete Luttinger liquid
theory for abelian quantum Hall states, taking into account 
the presence of gauge fields and
the effects of disorder and Coulomb interactions.
Building on the results obtained in [I] and [III], we have applied the
Chern-Simons flux attachment procedure to the theory of chiral edge
bosons in the integral quantum Hall regime. 
The resulting action has the well known $K$-matrix structure, but
possesses new properties and offers many novel insights due to the
microscopic nature of its derivation. 

We have shown that the Chern-Simons procedure is ill-defined for
systems with counter-flowing edge modes, unless there are long-ranged
electron-electron interactions present to stabilize the action for the
charged mode. Without long-ranged interactions, the Hamiltonian of
the charged mode is unbounded from below.
Our description for these systems is not plagued by
non-universalities (a problem that phenomenological theories have),
because our velocity matrix is diagonal simultaneously with $K$.
We also find no ambiguities concerning the definition of the Hall
conductance on the edge. The chiral anomaly provides an elegant way of
using Laughlin's gauge argument.

Our treatment of tunneling operators has shown that the CS procedure
affects the charge and statistics of tunneling particles. 
An electron operator outside the sample retains its unit charge after
the mapping, but inside the charge is mapped to $\nu/m$. 
The operators inside and outside are related by a T-duality
transformation that inverts the compactification radius ($K$-matrix);
thus we have found a new geometrical interpretation for this duality. 

Our analysis of spatially separated edge modes has revealed an even
richer duality structure in the spectrum of quasiparticles, where the
reversal inside/outside is generalized to a reversal of the order of
edge channels. Samples with positive $m$ and negative $m$ are jointly
described in this picture.

We have shown that in the limit of large length scales,
the theory of ``clean'', spatially separated
edges describes identical physics as the ``dirty'' $K$-matrix theory.
In the case of counter-flowing modes this happens in a quite
nontrivial way.

Following [III], we have derived an effective Luttinger liquid theory
for tunneling processes into the edge at filling fractions which do
not lie at plateau centers. The long range Coulomb
interactions between edge and bulk states result in a suppression of
the neutral modes, which directly yields a continuous tunneling exponent
$1/\nu$, in accordance with experiments.

We have made a short analysis of  transitions between fqH plateaus,
employing the 
$Q$-field theory for percolating fractional edges. 
The results
indicate that the critical aspects
of the plateau transitions are quite generally the same for both the
integral and fractional regime.
However, the Chern-Simons mapping is not uniquely
described by Sl(2,Z).

\vskip3mm

\noindent
{\bf ACKNOWLEDGEMENTS}\newline
This research was supported in part by the {\it Dutch Science
Foundation FOM} and by the {\it National Science Foundation} under
Grant No. {\it PHY94-07194}.


\vskip1cm
\scez\renewcommand{\theequation}{A\arabic{equation}}
\noindent
{\Large\bf Appendix A: Integrating out the Chern-Simons field}

\vskip0.4cm
\noindent
The remnant of the noninteracting part of the action 
(\ref{multform}) after $a_-$ has been integrated out
is given by
\bea
	-4\pi i \cdot S_{\rm remnant}^{\rm free} &=&
	s_m \sum_i\int_\infty\! \left[2D_-\qf_i\curl(\qy\vec D\qf_i)
	-\qy\vec D\qf_i\times\prt_-\vec D\qf_i\right]
	\nn\\
	&& -\fr{1}{2p}\int_\infty\!(2pm\qy+1)\vec a\times\prt_-\vec a
	-2m\int_\infty\!\qy\vec a\times\vec E
\label{remnant}
\eea
where $\qy$ is shorthand for $\qy(y)$ and $s_m$ stands for
$\sgn(m)$.
We now have to substitute (\ref{solconstr}) for $\vec a$,
\be
	\vec{a}=\fr{2p}{2pm\qy+1}s_m 
	[\qy {\textstyle\sum_i}\vec{D}\qf_i+\nabla\qO].
\label{vecaagain}
\ee
Let us first concentrate on all the terms that do not depend on the
arbitrary gauge $\qO$.
From the $\vec a\!\times\!\prt_-\vec a$ in (\ref{remnant}) we obtain a
term quadratic in $D_\mu\qf$, and the last term in (\ref{remnant})
produces a product of $A_\mu$ and $\vec D\qf$.
Together this is
\be
	-2p\sum_{ij}\int_\infty\!\fr{\qy^2}{2pm\qy+1}
	\vec D\qf_i\times\prt_-\vec D\qf_j
	-4p|m|\int_\infty\!\fr{\qy^2}{2pm\qy+1}
	{\textstyle\sum_i}\vec{D}\qf_i\times\vec E.
\label{noOmega}
\ee
The expression involving step functions can be simplified to
$\qy\nu/m$, since
$\prt_y[\qy^2/(2pm\qy\!+\!1)]\!=\!\prt_y[\qy\nu/m]$. 
Adding (\ref{noOmega}) to the first line of
(\ref{remnant}), we obtain 
\be
	-4\pi i\cdot S_{\qO=0}^{\rm free}=\nu\int\!\! A\wedge dA-\sum_{ij}
	[\qd_{ij}s_m -2p\fr{\nu}{m}]\oint(D_x\qf_i D_-\qf_j
	-\qf_i E_x).
\label{SqO=0}
\ee
Now we concentrate on the $\qO$. From $\vec a\!\times\!\prt_-\vec a$
we get a quadratic term in $\qO$,
\be
	-2p\int_\infty\!\fr{1}{2pm\qy+1}\nabla\qO\times\prt_-\nabla\qO.
\ee
Integrating by parts and using the relation 
$\prt_y[1/(2pm\qy\!+\!1)] \!=\! -2p\nu\qd(y)$ this can be rewritten as
\be
	-(2p)^2\nu\oint\prt_x\qO \; \prt_-\qO.
\label{quadraticqO}
\ee
Both terms in the second line of (\ref{remnant}) give rise to linear
terms in $\qO$; written together this is
\be
	-4p\;s_m \int_\infty\!\fr{\qy}{2pm\qy+1}\nabla\qO\times
	\nabla{\textstyle\sum_i}D_-\qf_i
	=
	4p\fr{\nu}{|m|}\oint\prt_x\qO \; {\textstyle\sum_i}D_-\qf_i,
\label{linearqO}
\ee
where we have integrated by parts and used
$\prt_y[\qy/(2pm\qy\!+\!1)] \!=\! \qd(y) \nu/m$.
The free action after the CS mapping is given by 
(\ref{SqO=0})+(\ref{quadraticqO})+(\ref{linearqO}). It is easily seen
that the terms involving $\qO$ can be completely absorbed into a
redefinition of the $\qf$ fields according to
\be
	\tilde\qf_i=\qf_i-2p\;s_m \qO.
\ee
The mapping of the interaction term is simple. We write the Coulomb
contribution in the form
\be
	-\fr{1}{8\pi^2}\int_\infty\! U(\vec x-\vec x')
	\curl [\qy{\textstyle\sum_i}\vec {\cal D}\qf_i](\vec x) \;
	\nabla'\!\times\! [\qy{\textstyle\sum_j}\vec {\cal D}\qf_j](\vec x')
\ee
and note that in this expression
$\qy \sum_i[\vec D\qf_i-\vec a]$, after substituting (\ref{vecaagain}),
is equivalent to 
$\qy \fr{\nu}{m}\sum_i[\vec D\qf_i-2p\;s_m \qO]
=\qy \fr{\nu}{m}\sum_i \vec D\tilde\qf_i$.
With this last ingredient, the mapping is complete and
yields the result (\ref{mappedch2}).


\vskip1cm
\scez\renewcommand{\theequation}{B\arabic{equation}}
\noindent
{\Large\bf Appendix B: Integrating out the CS field in the presence of
a tunneling term}

\vskip0.4cm
\noindent
Integration over the CS gauge fields in the presence of a tunneling
term proceeds along the same lines as in Appendix~A. The remnant
of the free action is now given by
\bea
	&& -4\pi i \cdot S_{\rm remnant}^{\rm free} =
	s_m \sum_i\int_\infty\! \left[2D_-\qf_i\curl(\qy\vec D\qf_i)
	-\qy\vec D\qf_i\times\prt_-\vec D\qf_i\right]
\label{remnant2}
	\\
	&& -\fr{1}{2p}\int_\infty\!(2pm\qy+1)\vec a\times\prt_-\vec a
	-2m\int_\infty\!\qy\vec a\times\vec E
	-4\pi\int_{\qt_1}^{\qt_2}\!\! d\qt\;(\prt_0\qf_i-i\vd a_x)
	(\qt,\vec x_0) \nn
\eea
and we have to substitute
\be
	\vec a(\vec x)=
	\fr{2p\;s_m }{2pm\qy(y)+1}\left[\qy(y)\sum_i\vec D\qf_i
	+\nabla\qO+s_m  L \nabla\arctg\fr{x-x_0}{y-y_0}
	\right]
\label{subst2}
\ee
with $L_n$ as defined in (\ref{defL}).
Eq. (\ref{subst2}) would suggest that one can obtain an action of
precisely the same form
as in Appendix~A from (\ref{remnant2}) (minus the tunneling term at
$\vec x_0$) by redefining $\qO$ according to 
$\qO' \!=\!\qO\!+\!s_m  L\arctg\fr{x-x_0}{y-y_0}$.
However, whereas $\qO$ satisfies $\curl\nabla\qO\!=\!0$, we have
$\curl\nabla\arctg\fr{x-x_0}{y-y_0}\!=\!-2\pi \qd(\vec x\!-\!\vec x_0)$.
This means that apart from the free part of (\ref{mappedch2}) 
(containing $\qO'$ instead of $\qO$) we get extra contributions at
$\vec x_0$.
The $\vec a\!\times\!\prt_-\vec a$ part of (\ref{remnant2}) gives rise
to the following
\be
	-2p\;s_m  \int_\infty\!
	\fr{1}{2pm\qy+1}L\nabla\arctg\fr{x-x_0}{y-y_0}\times\prt_-
	\left[s_m L \nabla\arctg\fr{x-x_0}{y-y_0}+2\qy\sum_i\nabla\qf_i
	+2\nabla\qO\right].
\ee
Integrating by parts and discarding all derivatives of step functions
(which yield $\oint$ terms that we have already accounted for), we
move all the derivatives in $[\cdots]$ to the left so that they can
operate on the $\arctg$. This gives
\be
	4\pi p \int_{\qt_1}^{\qt_2}\!\!d\qt\left[
	2\fr{\nu}{m}\qy\sum_i\prt_-\qf_i+2s_m (1-2p\nu\qy)\prt_- \{L
	\arctg\fr{x-x_0}{y-y_0}\}+2(1-2p\nu\qy)\prt_-\qO
	\right](\vec x_0).
\label{fromatimesa}
\ee
Notice the factor of 2 that has popped up in front of the $\arctg$
term. Its origin lies in the fact that $\prt_x$ and $\prt_y$ do not
commute when acting on $\arctg$. We have 
\be
	\prt_x\prt_y= \curl\nabla+\prt_y\prt_x
\ee
and the $\curl\nabla$ produces an extra delta function.
From the $\vec a\!\times\!\vec E$ part of (\ref{remnant2}) we get
\be
	2m\int_\infty\!\qy A_-\curl \vec a \longrightarrow
	-8\pi\; p\nu\qy\int_{\qt_1}^{\qt_2}\!\!d\qt\; A_-(\vec x_0).
\label{fromatimesE}
\ee
The $\int\!d\qt$ part of (\ref{remnant2}) gives
\be
	-4\pi\int_{\qt_1}^{\qt_2}\!\!\! d\qt\!\left\{\prt_0\qf_i
	-i\vd 2p\;s_m \left[\fr{\nu}{m}\qy\sum_j D_x\qf_j
	+(1-2p\nu\qy)\prt_x(\qO+s_m L\arctg\fr{x-x_0}{y-y_0})
	\right]\right\}
\label{fromdt}
\ee
Adding (\ref{fromatimesa}), (\ref{fromatimesE}) and (\ref{fromdt}) we
find 
\be
	S_{{\rm at }\;\vec x_0}^{\rm free}=-i\int_{\qt_1}^{\qt_2}\!\!\! d\qt
	\left\{\prt_0\tilde\qf_i(\vec x_0)
	-2p\fr{\nu}{m}\qy(y_0)\sum_j D_0\tilde\qf_j(\vec x_0)
	\right\}
\label{Satx0}
\ee
where we have defined
\be
	\tilde\qf_i(\vec x)=\qf_i(\vec x)
	-2ps_m\qO(\vec x)-2pL\arctg\fr{x-x_0}{y-y_0}.
\ee

Now we deal with the interaction term, in the same way as in
Appendix~A. Again we can write
\be
	\qy\sum_i(\vec D\qf_i-\vec a)\longrightarrow
	\qy\fr{\nu}{m}\sum_i\vec D\tilde\qf_i.
\ee
Only one subtlety is left: because of the presence of the $\arctg$ in 
$\tilde\qf$, the relation 
$\curl\vec D\tilde\qf_i \!=\! -B$ gets altered. We have
\be
	\curl\vec D\tilde\qf_i \!=\! -B'
	\hskip5mm ; \hskip5mm
	B'(\vec x)=\curl\vec A(\vec x)-2p\qd(\vec x-\vec x_0)
\ee
which reflects the fact that there is extra flux at $\vec x_0$.


\vskip1cm
\scez\renewcommand{\theequation}{C\arabic{equation}}
\noindent
{\Large\bf Appendix C: Tunneling exponent resulting from neutral modes}

\vskip0.4cm
\noindent
The result $S_{\rm out}$ (\ref{Soutside}) is obtained
from the expectation value
$\langle \exp-i\tilde\qf_k|_{\qt_1}^{\qt_2} \rangle$
as follows.
The field $\tilde\qf_k$ decomposes into charged and neutral
modes according to (\ref{basisneutral}). From the action (\ref{Sdiag})
we find the correlations for the diagonal modes,
\bea
	\langle\qG(\qt_2)\qG(\qt_1)\rangle=\fr{1}{\nu}\ln(\qt_2-\qt_1)
	& \hskip5mm ; \hskip5mm &
	\langle\qg_k(\qt_2)\qg_{k'}(\qt_1)\rangle
	=\qd_{kk'}\fr{k+1}{k}\ln(\qt_2-\qt_1)
\eea
and $\langle\qg_k\qG\rangle\!=\!0$.

Evaluation of 
$\langle \exp-i\tilde\qf_k|_{\qt_1}^{\qt_2} \rangle$
for arbitrary $k$
then yields the tunneling exponent
\bea
	S_{\rm out}&=&\frac{1}{\nu}+\frac{k-1}{k}+\sum_{a=k}^{|m|-1}
	\frac{1}{a(a+1)}
	=\frac{1}{\nu}+1-\frac{1}{k}+\sum_{a=k}^{|m|-1}
	\left(\frac{1}{a}-\frac{1}{a+1}\right) \nn\\
	& = &
	2p+1+\frac{1}{m}-\frac{1}{|m|}.
\eea


\end{document}